\documentclass{aa-files/aa}

\usepackage{graphicx}
\usepackage{xcolor}
\usepackage{txfonts}
\definecolor{xlinkcolor}{cmyk}{1,1,0,0}
\usepackage[colorlinks = true,
            linkcolor = xlinkcolor,
            urlcolor  = xlinkcolor,
            citecolor = xlinkcolor,
            anchorcolor = xlinkcolor]{hyperref}
\usepackage{fontawesome}
\usepackage{bm}

\usepackage[xindy]{glossaries}	

\newacronym{map}{MAP}{maximum \textit{a posteriori}}
\newacronym{dae}{DAE}{Denoising Auto-Encoder}
\newacronym{cnn}{CNN}{Convolutional Neural Network}
\newacronym{mse}{MSE}{mean-square-error}
\newacronym{snr}{SNR}{signal-to-noise ratio}
\newacronym{nfw}{NFW}{Navarro-Frenk-White}
\newacronym{2pcf}{2PCF}{two-point correlation function}
\newacronym{dsm}{DSM}{Denoising Score Matching}
\newacronym{rmse}{RMSE}{root-mean-square error}

\newcommand\yb{\boldsymbol{y}}
\newcommand\xb{\boldsymbol{x}}
\newcommand\zb{\boldsymbol{z}}
\newcommand\wb{\boldsymbol{w}}
\newcommand\nb{\boldsymbol{n}}
\newcommand\rb{\boldsymbol{r}}
\newcommand\mb{\boldsymbol{m}}

\newcommand\Sb{\boldsymbol{\mathrm{S}}}

\newcommand\Nb{\boldsymbol{\mathrm{N}}}

\newcommand\Pb{\boldsymbol{\mathrm{P}}}
\newcommand\Fb{\boldsymbol{\mathrm{F}}}
\newcommand\kp{\kappa}
\newcommand\g{\gamma}

\newcommand\score{\nabla\log p}
\newcommand\solarmass{M_\odot}
\newcommand{\hMpc}{h^{-1} \, \mathrm{Mpc}}

\newcommand\Rset{\mathbb{R}}

\newcommand{\nblink}[1]{\href{https://github.com/CosmoStat/jax-lensing/blob/master/papers/Remy2021/#1.ipynb}{\faFileCodeO}}
\newcommand{\github}{\href{https://github.com/CosmoStat/jax-lensing}{\faGithub}}

\begin{document} 

   \title{Probabilistic Mass Mapping with Neural Score Estimation}

   \author{B. Remy
          \inst{1}\fnmsep\thanks{Contact: benjamin.remy@cea.fr}
          \and
          F. Lanusse\inst{1}
          \and
          N. Jeffrey\inst{2,3}
          \and
          J. Liu\inst{4,5,6}
          \and J.-L. Starck\inst{1}
          \and K. Osato\inst{7,8}
          \and T. Schrabback\inst{9}}
          
   \institute{AIM, CEA, CNRS, Universit\'e Paris-Saclay, Universit\'e Paris Diderot, Sorbonne Paris Cit\'e, F-91191 Gif-sur-Yvette, France
         \and
             Laboratoire de Physique de l'Ecole Normale Sup\'erieure, ENS, Universit\'e PSL, CNRS, Sorbonne Universit\'e, Universit\'e de Paris, Paris, France
        \and University College London
             Gower St, London, UK 
        \and
             Berkeley Center for Cosmological Physics, University of California, 341 Campbell Hall, Berkeley, CA 94720, USA
             \and
             Lawrence Berkeley National Laboratory, 1 Cyclotron Road, Berkeley, CA 93720, USA
             \and
             Kavli IPMU (WPI), UTIAS, The University of Tokyo, Kashiwa, Chiba 277-8583, Japan
             \and
             Center for Gravitational Physics,
             Yukawa Institute for Theoretical Physics, Kyoto University,
             Kitashirakawa Oiwakecho, Sakyo-ku, Kyoto 606-8502, Japan
             \and
             LPENS, D\'epartement de Physique, \'Ecole Normale Sup\'erieure,
             Universit\'e PSL, CNRS, Sorbonne Universit\'e, Universit\'e de Paris,
             24 rue Lhomond, 75005 Paris, France
             \and
             Argelander-Institut f\"ur Astronomie, Universit\"at Bonn, Auf dem H\"ugel 71, D-53121 Bonn, Germany
             }

   \date{Received ...; accepted ...\\Report number: YITP-21-159}

 
  \abstract
   {
Weak lensing mass-mapping is a useful tool to access the full distribution of dark matter on the sky, but because of intrinsic galaxy ellipticies and finite fields/missing data, the recovery of dark matter maps constitutes a challenging ill-posed inverse problem. 
} 
   {We introduce a novel methodology allowing for efficient sampling of the high-dimensional Bayesian posterior of the weak lensing mass-mapping problem, and relying on simulations for defining a fully non-Gaussian prior. We aim to demonstrate the accuracy of the method on simulations, and then proceed to applying it to the mass reconstruction of the HST/ACS COSMOS field.}
   {The proposed methodology combines elements of Bayesian statistics, analytic theory, and a recent
   class of Deep Generative Models based on Neural Score Matching. This approach allows us to do the following: 1) Make full use of analytic cosmological theory to constrain the 2pt statistics of the solution. 2) Learn from cosmological simulations any differences between this analytic prior and full simulations. 3) Obtain samples from the full Bayesian posterior of the problem for robust Uncertainty Quantification.
   }
   { We demonstrate the method on the $\kappa$TNG simulations \citep{Osato2021} and find that the posterior mean significantly outperfoms previous methods (Kaiser-Squires, Wiener filter, Sparsity priors) both on root-mean-square error and in terms of the Pearson correlation. We further illustrate the interpretability of the recovered posterior by establishing a close correlation between posterior convergence values and SNR of clusters artificially introduced into a field. Finally, we apply the method to the reconstruction of the HST/ACS COSMOS field and yield the highest quality convergence map of this field to date.
   }
   {We find the proposed approach to be superior to previous algorithms, scalable, providing uncertainties, and using a fully non-Gaussian prior. All codes and data products associated with this paper are available at this link \github}
   {}


   \maketitle
%

\section{Introduction}

    The weak gravitational lensing effect provides a direct probe of the large scale matter distribution in the Universe. This lensing effect generates minute deformations of the apparent shapes of distant galaxies, a so-called shear, in the presence of massive structures along the line of sight. Due to its ability to directly probe the matter field, weak lensing is found at the heart of present and upcoming wide-field optical
    galaxy surveys including the ESA Euclid mission \citep{Laureijs2011}, the Vera C. Rubin Observatory Legacy Survey of Space and Time \citep{Ivezic2019},
    and the Roman Space Telescope \citep{Spergel2015}.
    
    While most of the cosmological analysis of weak lensing focuses on 2pt functions, the reconstruction
    of maps of the matter distribution opens up alternative ways to analyse data, including giving access to higher order statistics, such as the peak count statistic which has been applied to most existing weak lensing surveys~\citep{Liu2015,Liux2015,Kacprzak2016,Shan2018,Martinet2018,Harnois-Deraps2020}. In addition, novel higher-order statistics such as the wavelet peak counts and $\ell_1$-norm \citep{ajani2021}, the scattering transform statistics \citep{Cheng2020}, or neural summaries \citep[e.g.][]{Ribli2019, lfi_neural_2020} have recently been shown to be even more sensitive to cosmology.
    
    This process of reconstructing maps of the matter distribution from measured galaxy ellipticities is known as weak lensing mass-mapping. Because of noise and missing data, the weak lensing mass-mapping problem is ill-posed; in other words, the matter density field is not uniquely determined by the observed shear. This implies that any mass-mapping method will rely on different prior assumptions to regularize this problem and yield a different map estimate. The most standard approach is the Kaiser-Squires method \citep{Kaiser1993} which is based on a direct inversion of the lensing operator along with some amount of Gaussian smoothing. A number of other methods have since been proposed, using various types of regularisation schemes such as maximum entropy \citep{Marshall2001}, Gaussian prior (Wiener filter) \citep{simon2012,Horowitz2018}, Sparsity \citep{wlens:starck06,leonard2012,Lanusse2016,Price2018,sta:starck21}. These techniques usually yield a point estimate of a convergence map, and usually lack proper uncertainty quantification, which makes interpreting the resulting maps difficult. 
    
    A number of Bayesian methods have been proposed to recover a posterior probability estimate for the unknown mass map, but these are usually limited by restrictive prior assumptions. For instance, using hierarchical Bayesian modeling \citet{Alsing2015} proposed an approach for sampling mass-maps, but which relied on a Gaussian prior for the unknown map. More recently, \citet{Porqueres2021} extended this work and demonstrated mass-map posterior sampling with a non-linear prior defined by a forward gravity model, but limited to very low angular resolution. Other Bayesian posterior sampling approaches include \citet{Schneider2016} which relied on a Gaussian Process prior, \citet{Price2018} which proposed a proximal MCMC approach to accommodate non-differentiable sparsity priors, and \citet{Fiedorowicz2021} which relies on a log-normal prior.

    More recently, with the rise of deep learning, a number of methods relying on deep neural networks have been proposed to address the mass-mapping problem.
    The strength of these approaches is that they provide a practical way to leverage simulations as a prior to solve the mass-mapping problem. In particular, the \texttt{DeepMass} method \citep{Jeffrey2020} uses a U-net \citep{Ronneberger} to recover an estimate of the mean posterior convergence map, with a prior defined by a set of simulations. Besides, \citet{Shirasaki2021} have proposed a model based on a Generative Adversarial Network (GAN) \citep{Goodfellow2014} which is able to to denoise weak lensing mass maps.

    \bigskip
    In this paper, we propose a new approach to mass-mapping that combines elements of deep learning with Bayesian inference and provides a tractable way to sample from the full high-dimensional posterior distribution of convergence maps.
    
    The limiting factor in all Bayesian approaches mentioned above are the simplifications needed to achieve a tractable prior (which leads to Gaussian, log-Normal, sparse, or simplified hierarchical priors). These approximations are needed because the distribution of mass-maps is not tractable analytically, and only accessible as an \textit{implicit distribution}, i.e. a distribution without an explicit likelihood, and that can only be sampled from. Sampling from such an implicit simulation is otherwise known as running a simulator.

    Our approach \texttt{DeepPosterior}
    is based on learning a prior from samples drawn from such an implicit distribution (i.e. from simulated convergence maps), and use this prior for sampling the full Bayesian posterior.
    
    Instead of learning a full probability density function $p(x)$ over mass-maps with likelihood-based Deep Generative Models such Normalizing Flows \citep{Rezende2015a} or PixelCNNs \citep{Oord2016, Salimans2017}, following recent developments in the field of Diffusion-based models, we target instead the \textit{score function} $\frac{\partial \log p (x)}{\partial x}$. This score function is estimated using the Denoising Score Matching (DSM) technique \citep{Vincent2011}, which relies on training a neural network under a simple denoising task, and yet leads asymptotically to an unbiased estimate of the score function. 
    With this Neural Score Estimation in hand, we then demonstrate that one can sample from the posterior of the mass-mapping problem using an efficient gradient-based sampling techniques: Annealed Hamiltonian Monte-Carlo. Contrary to most similar Deep Learning approaches to solve inverse problems, this method is stable and scalable, we achieve one independent sample from the mass-mapping posterior for maps  of size $360 \times 360$ pixels in 10 GPU-minutes.
    
    We demonstrate our proposed methodology on simulations, using the high resolution $\kappa$TNG convergence maps \citep{Osato2021}, based on the IllustrisTNG simulations \citep{Nelson2019, Pillepich2018, Nelson2018, Springel2018, Naiman2018, Marinacci2018}. We compare the method against other standard methods (Kaiser-Squires, \texttt{DeepMass}, \texttt{GLIMPSE} \citep{Lanusse2016}, \texttt{MCALens} \citep{sta:starck21}) and find and improvement in terms of the pixel reconstruction error, the Pearson correlation and the convergence power spectrum. We also investigate the interpretation of the recovered posterior and demonstrate a direct correlation between signal to noise of structures like galaxy clusters and the posterior distribution.

    Following these validation tests on simulations, we apply the method to the reconstruction of the HST/COSMOS field \citep{Scoville2007} based on the shape catalog from \citet{Schrabback2010}. We obtain the highest quality mass map of the HST/COSMOS field, alongside uncertainty quantification. Our result improves over the previously published COSMOS map \citet{Massey2007DarkScaffolding} both from using a more recent shape catalog, and from our much improved methodology, revealing much finer structures and providing uncertainty quantification.
    
    The paper is organized as follows: After providing some general background on gravitational lensing in \autoref{sec:weaklensing}, and describing a unified view of mass-mapping and related works in \autoref{sec:massmapping}, we introduce in \autoref{sec:nsm} the methodology. We first describe how to build a prior from cosmological simulations and how to sample from the posterior distribution. Then in \autoref{sec:simulations}, we describe the simulations we used to train our model. In \autoref{sec:validation} we validate our method, showing improvement in point estimate reconstruction against other methods, and present a detection experiment using the posterior samples. Finally, in \autoref{sec:cosmos-recon}, we apply our method a real data, reconstructing a very high quality map of the HST/ACS COSMOS field.
    
\section{Weak Gravitational Lensing Formalism}
\label{sec:weaklensing}

In this section, we present an overview of weak gravitational lensing and mass-mapping.

Observed galaxy shapes are affected by the gravitational shearing effect that occurs in the presence of massive structures, acting as lenses, along the line of sight. This distortion can be described by a coordinate transformation \citep{Bartelmann2010} between unlensed coordinates $\beta$ and observed image coordinates $\theta$
\begin{equation}
    \beta = \theta - \nabla \psi(\theta),
\end{equation}
where $\psi$ is known as the lensing potential, and is sourced by the projected matter density on the sky.

To first order, the resulting distortions affecting galaxy images can be described in terms of a simple linear Jacobian matrix $\mathbf{A}$, known as the \textit{amplification matrix} :
\begin{equation}
    \beta = \mathbf{A} \theta = (1 - \kappa) \left( \begin{matrix} 1 - \gamma_1 & -\gamma_2 \\ -\gamma_2 & 1 + \gamma_1 \end{matrix} \right) \theta \;.
\end{equation}
In this expression, which only holds in the weak lensing regime (i.e. $\kappa \ll 1$), the \textit{convergence} $\kappa$, translates into an isotropic dilation of the source, while the \textit{shear} $\gamma$ causes anisotropic stretching of the image. 

This convergence $\kappa$ can be directly related to the projected mass density on the sky, which leads to typically using the denomination \textit{convergence map} or mass-map interchangeably. Indeed, considering a sample of lensing sources distributed in redshift according to some distribution $n(z)$, one can relate the convergence $\kappa$ to the 3D matter overdensity $\delta$ according to: 
\begin{equation} \label{eq:lensing-redshift}
    \kappa(\theta) = \frac{3H_0^2\Omega_\mathrm{m}}{2c^2}\int_0^{\chi_\text{lim}} d\chi \frac{q(\chi)}{a(\chi)}f_K(\chi) \delta(f_K(\chi)\theta, \chi),
\end{equation}
where $H_0$ is the Hubble constant, $c$ the speed of light, $q(\chi) = \int_\chi^{\infty} d\chi'~ n(\chi') \frac{f_K(\chi'-\chi)}{f_K(\chi')}$ is the lensing efficiency, $f_K$ is the comoving angular distance, $\delta$ the over density,  $a$ is the scale factor and $\Omega_\mathrm{m}$ is the matter density \citep{Kilbinger2015}.

While shear can be measured by the spatially coherent correlations it induces on galaxy shapes, convergence is typically not directly observable, as its magnification effect is much more difficult to disentangle from intrinsic galaxy sizes. Therefore, the problem of mass-mapping is generally to recover an estimate of the convergence $\kappa$ from measurements of the shear $\gamma$. This is made possible by the following equations that tie convergence and shear to the lensing potential:
\begin{equation}
    \kappa = \frac{1}{2} \Delta \psi \quad ; \quad \gamma_1 = \frac{1}{2} \left( \partial_1^2 \psi - \partial_2^2 \psi \right) \quad ; \quad \gamma_2 = \partial_1 \partial_2 \psi.
\end{equation}

Combining these equations, one can recover a minimum variance estimator for the convergence as:
\begin{equation}
    \kappa = \Delta^{-1} \left( (\partial_1^2 - \partial_2^2) \gamma_1 + 2  \partial_1 \partial_2 \gamma_2 \right),
\end{equation}
which constitutes the basis for the Kaiser-Squires reconstruction technique. This equation can be solved in practice most efficiently using a Fourier transform in the flat sky limit, or spherical harmonics transform in the spherical setting. 

The Fourier solution of the Kaiser-Squires estimator can be written as:
\begin{equation}
    \tilde{\kappa} = \frac{k_1^2 - k_2^2}{k^2} \tilde{\gamma}_1 + \frac{2 k_1 k_2}{k^2} \tilde{\gamma}_2,
    \label{eq:ks}
\end{equation}
where $k^2 = k_1^2 + k_2^2$. Note that the solution is not defined for $k=0$, which means that the mean of the convergence field cannot be directly constrained from shear, which is usually known as the \textit{mass-sheet degeneracy}.

One particularly remarkable property of the Kaiser-Squires estimator (\ref{eq:ks}) is that it defines a \textit{unitary operation}. In other words, let us introduce the linear operator $\mathbf{P}$ as:
\begin{equation} \label{eq:kappa-shear}
    \tilde{\kappa}_E +i \tilde{\kappa}_B =    \left( 
    \frac{k_1^2 - k_2^2}{k^2} + i\frac{2 k_1 k_2}{k^2}
    \right) \left(
    \tilde{\gamma}_1  + i \tilde{\gamma}_2
    \right) = \mathbf{P} \left(  \tilde{\gamma}_1  + i \tilde{\gamma}_2 \right),
\end{equation}
where $\tilde{\kappa} = \tilde{\kappa}_E + i \tilde{\kappa}_B$ and $\tilde{\gamma} = \tilde{\gamma}_1 + i \tilde{\gamma}_2$ are complex representations of the convergence E and B modes, and of the two shear components in Fourier space. Then the operator $\mathbf{P}$ verifies $\mathbf{P}^{\dagger} \mathbf{P} = \mathbf{I}_d$.

\section{A Unified View of Mass-Mapping} 
\label{sec:massmapping}

This mass-mapping problem can be reformulated as a probabilistic inference problem from a Bayesian perspective:

\begin{equation}
    p(\kp|\g) = \frac{p(\g|\kp)p(\kp)}{p(\g)},
\end{equation}
where the \textit{posterior} distribution $p(\kp|\g)$ models the probability of the signals $\kp$ conditioned on the observations $\g$, the \textit{likelihood} distribution $p(\g|\kp)$ encodes the forward process of the model in equation \autoref{eq:kappa-shear}, the \textit{prior} distribution $p(\kp)$ encodes the knowledge about the signal $\kp$, and the \textit{Bayesian evidence} $p(\g)$ is the marginal density of the observations. The evidence is a constant if we assume a given model, and will be ignored in the rest of this work as we do not consider Bayesian model comparison.

In our work, the forward process encoded in \autoref{eq:kappa-shear} returns a binned shear map $\gamma$, where each pixel value corresponds to the average shear in the pixel area and therefore takes as input a pixelized convergence map $\kp$. Working with real data, it is assumed that the measurement for each pixel is degraded by shape noise $n_s$ due to the finite average of galaxy intrinsic ellipticities in the bin. This noise is assumed to be white Gaussian, i.e. $n_s \sim \mathcal{N}(0, \Sigma_n)$, where $\Sigma_n$ is the noise covariance matrix of the shear map. Moreover, because of missing data due to survey measurement masks, we need to explicitly consider that there is no measurement in some pixel regions. In practice, we set a very high variance, such as $10^{10}$, for these pixels in the covariance matrix $\Sigma_n$. More specific information on the strategy we follow to emulate the COSMOS shape catalog can be found in \autoref{sec:emulated-cosmos-data}.

Thus, the log-likelihood takes the following form:
\begin{equation} \label{eq:likelihood}
    \log p(\g|\kp) = - \frac{1}{2}(\gamma - \Fb^*\Pb \Fb \kp)^{\dagger}\Sigma_n^{-1}(\gamma - \Fb^*\Pb \Fb \kp) + constant,
\end{equation}
where $\Fb$ and $\Fb^*$ are respectively direct and inverse Fourier transform.

All existing mass-mapping techniques can be understood under the lens of this Bayesian formulation and will generally mostly differ in their choice of prior, and in the specific algorithm used to recover a point estimate of the convergence map. As in practice this problem is ill-posed due to noise corruption and missing data in \autoref{eq:kappa-shear}, the posterior $p(\kp|\g)$ can be both wide and heavily prior dependent, which explains why all these different techniques yield different answers. Below  we describe several methods we use in this paper for comparison.

\subsection{Kaiser-Squires Reconstruction}

The Kaiser-Squires method \citep{Kaiser1993} can be seen as a simple maximum likelihood estimate (MLE) of the convergence map, typically followed by a certain amount of Gaussian smoothing.
\begin{align}
    \check{\kappa}_{ks} &= \arg \min_\kappa \parallel \gamma - \Fb^*\Pb \Fb \kappa \parallel_2^2 = (\Fb^*\Pb \Fb)^\dagger \gamma \label{eq:ks2}\\
    \kappa_{ks} &= s \ast \check{\kappa}_{ks}
\end{align}
where $s$ is a Gaussian smoothing kernel of a given scale, and $\mathbf{P}^\dagger$ is a pseudo-inverse of the operator $\mathbf{P}$ typically achieved by direct Fourier inversion. While this method is the fastest, it does not take into account masks and leads to leakage between E and B modes of the convergence field. For Kaiser-Squires, the heteroscedasticity does not impact the solution, whereas it can do for certain extensions such as the Generalized Kaiser-Squires method (GKS) \citet{sta:starck21} (Appendix B.1), in which an iterative approach with little regularization take the mask and noise heterescedasticity into account.

\subsection{Wiener Filter} \label{wiener-filter}

The Wiener filter approach assumes a Gaussian random field prior on $\kp$ and takes advantage of the fact that the power spectrum of the convergence can be analytically predicted from cosmological models, and accurately describes the field on large scales. 
This prior on the convergence can be expressed as a Gaussian distribution with a diagonal covariance matrix $\Sb$ in Fourier space:
\begin{equation} \label{eq:gaussian-prior}
    p_\text{\tiny Gaussian}(\kp) = \frac{1}{ \sqrt{ \det 2 \pi \mathbf{S}}} \exp \left( -\frac{1}{2} \tilde{\kp}^\dagger \Sb^{-1} \tilde{\kp} \right),
\end{equation}
where $\Sb$ is the convergence power spectrum.

The solution of the inverse problem can be formulated as:
\begin{equation}
    \hat{\kappa}_\mathrm{wiener} = \arg \min_\kappa \parallel \Sigma^{-1/2} (\gamma - \Fb^*\Pb \Fb \kappa) \parallel_2^2 + \log p_\text{\tiny Gaussian}(\kp) \;.
\end{equation}
This Wiener solution corresponds to the \textit{maximum a posteriori} (MAP) solution under this Gaussian prior, and also matches the mean of the Gaussian posterior. An appealing property of this estimator is that the solution can be easily recovered analytically in the case where the noise is homoscedastic, as both signal and noise covariance matrices become diagonal in Fourier space. The Wiener Filter reconstruction \citep{Lahav1994, Zaroubi1995} is given in Fourier space by:
\begin{equation}
    \tilde{\kp} = \Sb\mathbf{P}^\dagger\left[ \mathbf{P} \Sb \mathbf{P}^\dagger + \Nb \right]^{-1}\tilde{\g},
\end{equation}
where $\Sb$ and $\Nb$ are respectively the signal and noise covariance matrix in Fourier space.

In more complex cases, where the noise covariance is not diagonal in Fourier space (for instance because of a mask in pixel space), the solution can still be recovered efficiently by optimization, using the proximal method \citep{bobin2012wiener,sta:starck21}, or its related messenger field alternative \citep{elsner2013}. One can also draw samples from the Wiener posterior with the messenger field algorithm \citep{starck:jeffrey18}.

\subsection{Sparse priors} \label{sparse-prior}
Convergence maps contain non-Gaussian features, that are not well recovered with the methods described above. Several mass-mapping algorithms have been proposed, relying on a wavelet sparsity prior \citep{wlens:starck06,leonard2012,Lanusse2016,Price2018,sta:starck21}, which can be formulated as:
\begin{equation}
    \log p(x) = \parallel \mathbf{\Phi}^t x \parallel_p
\end{equation}
with $p < 2$, and where $\mathbf{\Phi}$ is a wavelet dictionary and $\parallel . \parallel_p$ is a sparsity promoting $\ell_p$ norm. 

The convergence map is the solution of the following sparse recovery optimization problem:
\begin{equation}
\hat{\kappa} = \arg\min_\kappa \parallel \Sigma^{-1/2}(\gamma - \Fb^*\Pb \Fb \kappa) \parallel_2^2 + \lambda \parallel \mathbf{\Phi}^t \kappa \parallel_p,
\end{equation}
where $\lambda$ is the regularisation parameter, weighting the sparse regularisation constraint.
The \texttt{GLIMPSE} method \citep{Lanusse2016}, allows in addition to take into account masks, non-uniform noise, flexion data if they are available, and also does not require to transform the shear catalog on pixelized map.

\subsection{\texttt{DeepMass}} \label{deepmass}

While all the priors described above have closed-form expressions, it is also possible to design a mass-mapping method where the prior is defined implicitly.
The first dark matter map reconstruction from weak lensing observational data using deep learning was shown in \cite{Jeffrey2020}. \texttt{DeepMass} is a Convolutional Neural Network (CNN) trained on pairs of simulated pixelized shear and convergence maps. One can show that under some assumptions, a deep learning model can estimate the mean of the posterior distribution $p(\kp|\g)$. In a nutshell, the network needs to be trained to minimize the the mean-squared-error (MSE) of the output convergence $\kp$ and the training convergence and shear maps must be drawn respectrively from the prior $p(\kp)$ and likelihood distribution $p(\g|\kp)$.

While the Wiener Filter assumes a Gaussian prior over the convergence, simulated training data for \texttt{DeepMass} are drawn from the "true" prior $p(\kp)$ and thus improve the accuracy of the reconstruction. \texttt{DeepMass} is therefore able to recover the non-linear structures of the convergence better than the Wiener filter and to reduce the MSE of the reconstruction.

Even if \texttt{DeepMass} reconstructs high quality convergence maps, \texttt{DeepMass} alone \textit{only provides the mean posterior and cannot quantify the uncertainties of the reconstruction}.
Furthermore, as any direct inversion method based on neural networks, the likelihood \autoref{eq:likelihood} is learned implicitly by the model, and does not explicitly constrains the solution at inference time. Meaning that although we have not found in our experiments obvious failures, the CNN \textit{may} in theory fail in ways that would lead to a map not actually consistent with observations, for instance creating spurious artifacts, or missing structures present in the true map. 
Another side effect of implicitly learning the likelihood during training is that the model is trained for a specific survey configuration, and retraining is required if either the mask or the noise is different.

In this work, we propose a new approach which estimates the full posterior distribution $p(\kp|\g)$,  being able to not only  recover the posterior mean, but also to quantify the uncertainties of the reconstruction. In addition, in our method the likelihood is explicit, meaning that it does not require a retraining for a new survey configuration.

\section{Primer on Neural Score Estimation and Sampling} \label{sec:nse}
In this section, we review the technical aspects of the machine learning methodology we will employ in the mass-mapping problem. We begin by detailing how the score function can be estimated for an implicit distribution. We then describe our strategy to sample any distribution from the knowledge of its score.

There are many classes of generative models in the machine learning litterature, such as GANs \citep{Goodfellow2014}, VAEs \citep{Kingma2014}, Normalizing Flows \citep{Rezende2015a}, or EBMs \citep{LecunEBM}.

All of these methods aim at modeling the probability distribution underlying some data, but not in the same way. On one hand, GANs or VAEs learn implicitly the probability distribution, which do not fit into our framework because we want to leverage the closed-form expression of the likelihood distribution. On the other hand, Normalizing Flows and EBMs can learn explicit forms of probability distribution, but do not scale well in dimension. Therefore we resort to a recent and promising class of generative models introduced in \citet{Song2019} based on learning an explicit model of the gradient log-probability distribution, also known as the \textit{score} function, which is all we need in our framework to build the posterior distribution as described in \autoref{sec:posterior-sampling}.

\subsection{Denoising Score Matching}\label{DSM}

\begin{figure}
    \centering
    \includegraphics[width=\columnwidth]{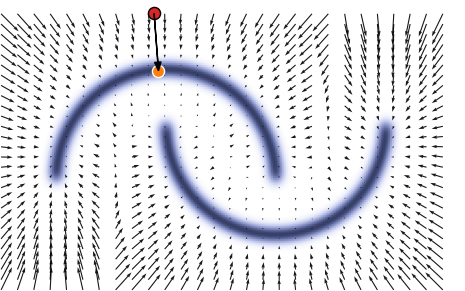}
    \caption{Illustration of the score field (represented by the vector field) on the simple two-dimensional "two moons" distribution. The score field points in the direction of regions of higher probability. A noisy sample from the distribution is shown in \textit{red}, while the denoised sample computed using equation \autoref{eq:denoiser} is shown in \textit{orange}. This illustrates how the variance multiplied by the score function,
    represented by the \textit{black} arrow,  can project the sample onto the high density of the distribution. \nblink{IntroductionToScoreMatching}}
    \label{fig:score_intro}
\end{figure}

As originally identified by \cite{Vincent2011} and \cite{Alain2013}, the gradient w.r.t. the data of a log distribution, which is called the \textit{score} function, can be modeled using a Denoising Auto-Encoder (DAE). This method is also known as Denoising Score Matching (DSM).

Let us introduce an auto-encoding function $\rb_\theta: \Rset^{N} \times \Rset \rightarrow \Rset^{N}, (\xb', \sigma)\mapsto \tilde{\xb}$, where $N$ is the signal dimension, trained to reconstruct a true signal $\xb$ following the probability distribution $p$, given a noisy version $\xb' = \xb + \nb$, with $\nb \sim \mathcal{N}(0, \sigma^2 \boldsymbol{I}_N)$, under an $\ell_2$ loss. In this case, the function $\rb_\theta$ is called a denoiser, parametrized by a noise level $\sigma$ and is trained minimizing the following criterion:
\begin{equation} \label{eq:loss1}
     \mathcal{L}_{\mathrm{DAE}} = \mathbb{E}_{\xb^\prime \sim p_{\sigma^2}}\left[ \|\xb - \rb_\theta(\xb', \sigma) \|_2^2 \right].
\end{equation}
where $p_{\sigma^2} = p * \mathcal{N}(0, \sigma^2)$, i.e. corresponds to the data distribution we want to model, convolved by a multivariate Gaussian with a diagonal covariance matrix $\sigma^2 \boldsymbol{I}_N$.

\cite{Alain2013} showed that in this setting, an optimal denoiser $\rb^\star$ would then be achieved for:
 \begin{equation} \label{eq:denoiser}
     \boldsymbol{r}^\star(\boldsymbol{x}, \sigma) = \boldsymbol{x} + \sigma^2 \nabla_{\boldsymbol{x}} \log p_{\sigma^2}(\boldsymbol{x}) + o(\sigma^2), ~~\mathrm{as} ~~\sigma^2 \rightarrow 0.
 \end{equation}
 \autoref{fig:score_intro} illustrates how one can denoise a 2-dimensional point knowing the noise level and the score function. We can clearly see that the score multiplied by the variance 
gives the right shift between the noisy point (in red) and the high density region of the distribution. Applying this shift to the noisy sample gives the projected sample in orange.

Rewriting this equation gives an estimator of the score function:
\begin{equation} \label{eq:score}
    \nabla_{\xb} \log p_{\sigma^2}(\xb) = \frac{\rb^\star(\xb, \sigma)- \xb}{\sigma^2} + o(1), ~~\mathrm{as} ~~\sigma^2 \rightarrow 0.
\end{equation}
The optimal denoiser is thus related to the score we wish to learn, and exactly matches the score of the probability distribution underlying data when the noise variance $\sigma^2$ goes to zero. 

Following the loss function design in \citet{Lim2020}, the optimization objective to train the denoiser is:
\begin{equation}
    \mathcal{L}_\text{AR-DAE} = \underset{\begin{subarray}{c}
  \xb \sim P \\
  \boldsymbol{u} \sim \mathcal{N}(0, I) \\
  \sigma \sim \mathcal{N}(0, s^2)
  \end{subarray}}{\mathbb{E}} \left[ \parallel  \boldsymbol{u} + \sigma \boldsymbol{r}_{\theta}(\boldsymbol{x} + \sigma \boldsymbol{u}, \sigma) \parallel_2^2  \right], \label{eq:dsn}
\end{equation}
where $\boldsymbol{u}$ is sampled from a standard multivariate Gaussian, $r_\theta$ is a denoiser parametrized by neural network weights $\theta$, optimized to estimate $r^*$. The modifications between \autoref{eq:loss1} and \autoref{eq:dsn}
resolve some numerical instabilities and reduce the error of approximation of the precedent objective.
In particular, the learned denoiser $\rb_\theta$ now directly models the score function, without the need to divide by the noise level anymore, i.e. $\rb_\theta(\xb, \sigma^2) = \nabla_{\xb} \log p_{\sigma^2}(\xb)$, as $\sigma^2 \rightarrow 0$. Moreover rescaling by $\sigma$ and decoupling the noise level from an isotropic Gaussian noise prevents the gradient from vanishing.

To summarize, DSM provides a tractable way of estimating a score function, from only having access to samples from an implicit distribution. We will now be able to use this score estimate for sampling from said distribution, as detailed in the next section.

\subsection{Score-based sampling} 
\label{sec:sampling-method}

Given the score function $\score (\xb)$ (either directly available, or learned by DSM), it is possible to sample from the  distribution $p(\xb)$ by Markov Chain Monte Carlo (MCMC). Indeed, the Langevin Dynamics (LD) or Hamiltonian Monte Carlo (HMC) updates, described in \citet{Neal2011} and \citet{Betancourt2017}, only require the evaluation of the score function $\score(\xb)$. For instance, the HMC update, based on the leapfrog integrator, requires the evaluation of the score function only:
\begin{equation}
    \begin{array}{lcl} \mb_{t+\frac{\alpha}{2}} & = & \mb_t + \dfrac{\alpha}{2} \score(\xb_t) \\ 
    \xb_{t+\alpha} & = & \xb_t + \alpha \boldsymbol{\mathrm{M}}^{-1} \mb_{t+\frac{\alpha}{2}} \\
    \mb_{t+\alpha} & = & \mb_{t+\frac{\alpha}{2}} + \dfrac{\alpha}{2} \score (\xb_{t+\alpha})
    \end{array}
    \label{eq:hmc}
\end{equation}
where $\alpha$ is the step size, $\boldsymbol{m}$ is the auxiliary momentum and $\boldsymbol{\mathrm{M}}$ is a preconditioning matrix that could take into account the space metric, but in our case the identity matrix. 

Following this procedure, HMC is supposed to sample from $p(\xb)$, but as explained in \cite{Betancourt2017}, the discretization induces a small error that will bias the resulting transition and requires a correction. In order to correct this bias, every sample is considered as a Metropolis-Hastings (MH) \citep{Metropolis1953, Hastings1970} proposal and is accepted or rejected according to an acceptance probability. This acceptance probability is designed from the Hamiltonian transition and is also only score-dependent as we show in \autoref{sec:MH}.

However, in most, if not all but the most trivial, cases sampling in high-dimension by Langevin or Hamiltonian dynamics is made very difficult by the fact that the distribution manifold is never Euclidean, meaning that assuming a diagonal noise covariance for LD, or diagonal momentum matrix for HMC, leads to extremely inefficient sampling.

To take a concrete example, let us consider the case of distribution of handwritten digits in the MNIST dataset, and imagine a Langevin Dynamics chain exploring this distribution. To transition from a 1 to a 7 for instance, the chain running in pixel space will try to add some white Gaussian noise at each update. However, MNIST digits are binary, so any addition of noise is bound to kick the chain out of the data distribution, and be rejected in a Metropolis-Hastings step. Only the smallest step sizes $\epsilon$ (which also tunes the amount of noise applied on the chain) will have a non-zero chance of acceptance, meaning the chain is practically never moving.

To summarize, having  access to the score function is all we need to sample from a distribution, but this remains difficult for non-trivial high dimensional distributions. We describe in the next subsection a strategy to circumvent this issue.

\subsection{Efficient sampling in high-dimensions with annealing}

Various approaches have been suggested to increase the sampling efficiency of HMC and LD for complex distributions, which generally aim at reframing the chain in a space closer to the intrinsic manifold of the distribution rather than pixel space. For instance, \citet{Girolami2011} exploits the Riemannian geometry of parameters space to define a Metropolis-Hastings \citep{Metropolis1953, Hastings1970} proposal, enabling high efficiency sampling in high dimensions.

In our work, we follow another direction which recently gained momentum from the literature on Denoising Diffusion Models \citep{Sohl-Dickstein2015DeepThermodynamics, Song2019, Ho2020DenoisingModels}. Instead of trying to follow the non-Euclidean latent manifold of the distribution, why not transforming this distribution so that it becomes easier to travel in pixel-space? This can be done very simply by convolving the data distribution with noise.

\begin{figure}[h!]
    \centering
    \includegraphics[width=\linewidth]{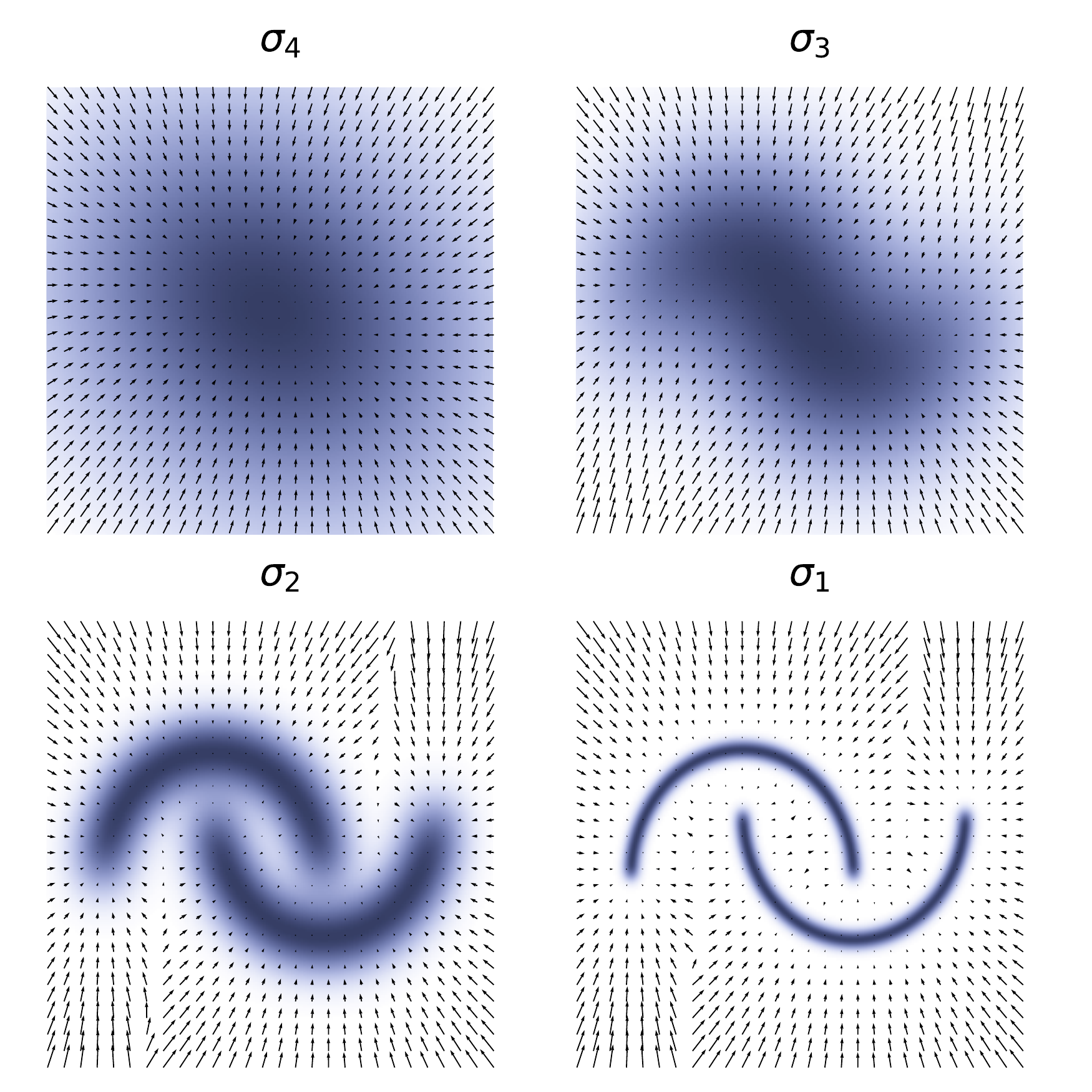}
    \caption{Example of annealing on the two moons distribution. The distribution is convolved with a multivariate Gaussian of variance $\sigma$. This variance is decreasing over the sampling procedure so that the MCMC is always over high density regions, i.e. $\sigma_4 > \sigma_3 > \sigma_2 > \sigma_1$. The chain is initialized from a wide multivariate Gaussian, as on the top left panel, and converges to the data distribution, as on the bottom right panel. In blue is the density of the convolved distribution and the arrows represent its \textit{score function}.}
    \label{fig:annealing}
\end{figure}

To go back to our MNIST thought experiment, if we convolve our binary handwritten digits with Gaussian noise, now it becomes very easy for a LD chain to move in pixel space, as the noise added by the chain can be made to match that present in the distribution. The higher the noise, the largest pixel-wise transition will be possible, and the faster the chain can transition from one digit to the other. Another view of this effect is that given enough noise, all of the distinct modes of a distribution will begin to merge with each other. \autoref{fig:annealing} illustrates the same two-moons distribution convolved with varying amount of Gaussian noise. At high noise values (top left corner), the distribution becomes close to a diagonal Gaussian and thus extremely easy to sample to explore all possible regions of the distribution.

We will call \textit{temperature T} the variance $\sigma^2$ of this Gaussian kernel convolved with the target distribution.

\bigskip

Following our preliminary work \citep{Remy2020}, we adopt in this paper a two-step sampling procedure, based on an \textit{annealed} version the Hamiltonian Monte Carlo (HMC) \citep{Neal2011, Betancourt2017}, similar to the annealed Langevin Dynamics proposed by \cite{Song2019}. 

In the first step of our sampling procedure, we initialize a chain using white Gaussian noise with a high temperature $\sigma_T^2$. Leveraging the conditional noise property of the score function described in section \ref{DSM} we have direct access to the score function of the convolved distribution $\score_{\sigma_T^2}$, which we can use in a score-based HMC. We then let the chain evolve under Hamiltonian dynamics, and progressively lower the temperature  $\sigma^2$ of the conditional score. Each time the temperature $\sigma^2$ is decreased, the MCMC chain thermalizes to the new temperature in a few HMC steps. As described in \citet{Song2019a}, small enough steps are important to ensure non-zero transition probabilities between adjacent temperature values. Once the chain has reached sufficiently low temperature (ideally $\sigma^2 =0$) we stop the chain and \textit{only retrieve the last sample}. Multiple independent samples are obtained by running multiple independent annealed HMC chains in parallel. One must not think that having one chain per sample makes the process much longer than having multiple sample per chain. It is indeed much more efficient, because in practice it is very long to ensure sample independence within the same chain.

In practice however, it has proven to be difficult to anneal chains all the way to zero temperature. We find that at low temperatures, the chains do not properly thermalize at each step and residual noise remains in our samples. This was also observed by other authors using annealed Langevin Dynamics instead of HMC \citep{Jolicoeur-Martineau2020, Song2019}. In the application in this paper, we can reach $\sigma=10^{-3}$ by fine-tuning our annealing scheme.

Therefore, in a second step of our sampling procedure, we propose a strategy to transport the annealed HMC samples obtained in step 1 at finite temperature all the way to $\sigma^2 = 0$. To achieve this, we use remarkable results from \citet{Song2020a} which establish a parallel between denoising diffusion models (generative models based on random walks) and Stochastic Differential Equations. We refer the interested reader to \autoref{sec:ode} for more mathematical details and derivations, but the main result from that paper that we use here is the following Ordinary Differential Equation:
\begin{equation}\label{eq:ve-ode}
    d\xb = -\frac{1}{2}\sqrt{\frac{d}{dt}\Big(\sigma^2(t)\Big)}\nabla_{\xb}\log p_t(\xb)dt  \;,
\end{equation}
where $\sigma^2(t)$ is the increasing variance schedule of the Gaussian distribution we convolve the data distribution with. In our particular implementation, we used a linear schedule $t \mapsto \sigma^2(t) = t$.
This ODE describes a deterministic process $\{\xb(t)\}_{t=0}^T$ indexed by a continuous time variable $t\in[0,T]$, such that $\xb(0)\sim p_0$ and $\xb(T)\sim p_T$, where $p_0$ denotes the data distribution and $p_T$ the convolution between the data distribution and a multivariate Gaussian of variance $T$. This ODE can be solved by any black-box ODE solver provided that the convolved score is available. In particular, it means that if we start the ODE at a point in $p_t$, where $t$ is the intermediate temperature at which we have stopped our HMC chains, we can transport that point to $p_0$. This procedure very effectively removes residual noise while making sure we reach a point in the target distribution at zero temperature.

\medskip

We summarize our full sampling procedure as follows:

\begin{enumerate}
    \item Initialize with white Gaussian noise $x_\text{init} \sim \mathcal{N}(0, T\cdot \bm{\mathrm{I}}_d)$
    \item Sample the posterior distribution with the \textit{annealed HMC} algorithm which requires at each step to:
    \begin{enumerate}
        \item evaluate the convolved score $\score_\sigma (\g|\kp)$ using \autoref{eq:score},
        \item compute the MH proposal using \autoref{eq:hmc} and accept it according to \autoref{eq:score-MH},
        \item anneal if there is enough samples for a given temperatures.
    \end{enumerate}
    \item Project the last sample on the posterior distribution with the \textit{ODE} described in \autoref{eq:ve-ode}.
\end{enumerate}

\section{Mass-Mapping by Neural Score Estimation} \label{sec:nsm}
Having laid out the machine learning notions needed to implement our method, we now describe how to apply these techniques to the weak lensing mass-mapping problem. Our strategy will be to use Denoising Score Matching to learn a high fidelity prior over convergence maps, and use this prior in combination with the analytic likelihood function \autoref{eq:likelihood} to access the full posterior of the mass-mapping problem by annealed HMC sampling.

\begin{figure*}[h]
    \centering

    \includegraphics[width=\textwidth]{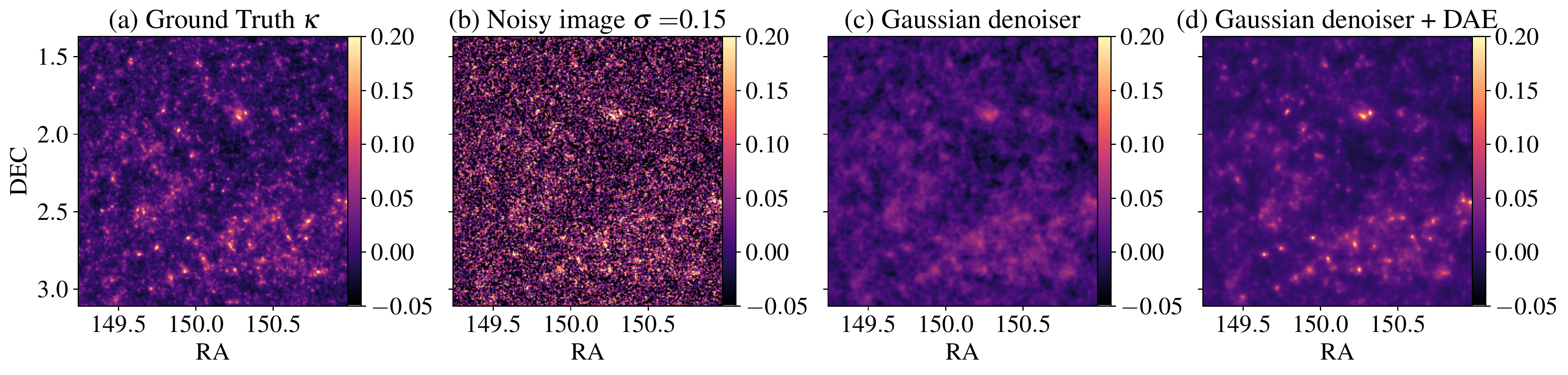}

    \caption{Denoising of a $100\times100$ arcmin$^2$ simulated convergence map (a) which was corrupted with a noise level of $\sigma=0.15$ per pixel (b). The Gaussian denoiser (c), computed with equations \ref{eq:denoiser} and \ref{eq:gaussian-prior}, retrieves large scales of the convergence map. Adding the DAE residuals to the Gaussian denoiser (d) allows us to recover both large and small scales of the $\kp$ map. This illustrates the Gaussian (analytic) and non-Gaussian (learned) parts of our hybrid score function \autoref{eq:score}. \nblink{HybridDenoiser}
    }
    \label{fig:denoiser}
\end{figure*}

\subsection{Hybrid analytic and generative modeling of the prior} \label{hybrid prior}

One of the main limiting factors of previous mass-mapping approaches is the limited complexity of the prior used in the inversion problem (e.g. Gaussian, or sparse). Our first step towards our mass-mapping technique is to build a prior that can fully capture the non-Gaussian nature of convergence maps. To achieve this goal, we propose an hybrid prior relying on a Gaussian analytic model, complemented by a Denoising Score Matching approach to learn on simulations the delta between this Gaussian prior and fully non-Gaussian convergence map distribution, accessible through simulations.

We indeed know that analytic cosmological models provide a reliable model of the 2pt functions of the convergence field, and can thus be used to define a Gaussian prior which enjoys a closed-form expression (\autoref{eq:gaussian-prior}), as discussed in \autoref{wiener-filter} in the context of the Wiener filter.  Moreover, because of its tractable expression we directly have access to its score function $\score_\text{Gaussian}(\kp)$. This Gaussian prior is particularly adapted on large scales, where the matter distribution is well modelled by a Gaussian Random Field, but does not capture the significantly non-Gaussian nature of the convergence field on small scales. 

On the other hand, we do have access to physical models that can capture the full statistics of the convergence field, in the form of numerical simulations. In this case however, the simulator gives us access to an \textit{implicit distribution} i.e. we can only draw samples from the distribution, but we do not have access to the likelihood of a given sample under the model. We cannot therefore directly use black-box simulators as priors for sampling the Bayesian posterior. This is where we can leverage deep generative models, to turn samples from an implicit distribution into a model with a tractable likelihood that can be used for inference. 

In this work, we propose to use Denoising Score Matching for this task, but we also want to capitalize as much as possible on the analytic Gaussian prior. Consequently, we aim to use the neural network to model the higher order moments of the convergence field that the Gaussian prior cannot capture. In a DSM framework, it is straightforward to implement such a network in practice by feeding the neural network denoiser with (1) the noisy input image (2) the noise level and (3) the Gaussian score. The full prior $p(\kp)$ is then computed as:

\begin{equation} \label{eq:full-prior}
    \score(\kp) = \score_\text{\tiny Gaussian}(\kp) + \rb_\theta\left( \kp, \score_\text{\tiny Gaussian}(\kp) \right), 
\end{equation}
where $\rb_\theta$ is the DAE trained to model only the residuals from the Gaussian prior. During learning, a first step of denoising is performed using $\score_\text{\tiny Gaussian}(\kp)$, then the neural network $r_\theta$ is optimized to denoise from this Gaussian guess, hence referred as the residual score.

Just like for 
conventional Denoising Score Matching, we can train the model on simulations by adapting \autoref{eq:dsn} to Residual Denoising Score Matching:

\begin{multline}
  \mathcal{L}_\text{RDSM} = \underset{\begin{subarray}{c}
  \boldsymbol{u} \sim \mathcal{N}(0, I) \\
  \sigma \sim \mathcal{N}(0, s^2)
  \end{subarray}}{\mathbb{E}} \parallel \boldsymbol{u} + \sigma *\big[ \boldsymbol{r}(\boldsymbol{x} + \sigma*\boldsymbol{u}, \sigma, \nabla \log p_\text{G}(\boldsymbol{x}+ \sigma*\boldsymbol{u})) \\ + \nabla \log p_\text{G}(\boldsymbol{x}+ \sigma*\boldsymbol{u}) \big] \parallel^2 \label{eq:residual_dsn}
\end{multline}
where $p_\text{G}$ is the Gaussian prior.

\autoref{fig:denoiser} provides an illustration of this hybrid prior. From the noisy input map (b), we first compute the Gaussian score function. Panel (c) shows the Gaussian-denoiser estimate computed with \autoref{eq:denoiser} from the Gaussian score function. Then, feeding the DAE with the noisy input, the input noise level and the Gaussian score function, we get the full score function. Panel (d) shows the complete denoising estimate computed with \autoref{eq:denoiser} again from the score function. \autoref{fig:denoiser} exemplifies the ability of the neural network to capture complementary scales to the Gaussian prior.

Building the prior with this hybrid design reduces the complexity of the modelling problem, and limits the reach of the neural network model by only modeling a correction to the analytical prior.

\subsection{Neural Score Estimator architecture for mass-mapping}

The Denoising Score Matching approach described in \autoref{DSM} was so far completely generic and agnostic of the actual implementation of the parametric denoiser $r_{\theta}$. We describe here the concrete neural network architecture we will use throughout this work, fine-tuned to image data of size 360$\times$360 pixels.

Following \autoref{eq:score}, we need to train a DAE in order to model the prior score function. We used a U-net, inspired from \cite{Ronneberger}, with ResNet building blocks of convolutions from \citet{He2016a} followed by Batch Normalization. We also used Spectral Normalization, using power iteration method on the neural network weights \cite{Gouk2018a}, to improve the regularity of the network in out-of-distribution regions, where the score is not constrained by the data. Indeed, as discussed in \autoref{sec:unet}, regularizing the spectral norm lowers the Lipschitz constant of the network and thus enforces the score field to be aligned for close inputs. Note that the network is noise conditional, which means that it takes as inputs both the image to denoise and the noise level. The noise level was drawn from a normal distribution of standard deviation $\sigma=0.2$. Our U-net denoiser was trained using the Adam optimizer \citep{Kingma2014} starting with a learning rate of $10^{-4}$ decreasing up to $10^{-7}$ at the end. We used a batch size of 32 to train the neural network for $40000$ steps.

More details on the architecture, training, and regularisation are given in \autoref{sec:unet}.

\subsection{Sampling from the posterior} \label{sec:posterior-sampling}

The prior learned with DSM is naturally defined as a convolved version of the data distribution. Besides, in order to use annealing on the posterior distribution, it is necessary to convolve also the likelihood.
Under the assumption that the likelihood is Gaussian, convolving with a multivariate Gaussian on $\xb$ gives the following expression:

\begin{equation} \label{eq:likelihood_conv}
    \log p_{\sigma_L^2}(\yb|\xb) = -\frac{\|\yb - \Pb\xb\|_2^2}{2(\sigma_n^2 + \sigma^2)} + constant ,
\end{equation}
where $\sigma_L^2$ is the variance of the convolved likelihood, $\sigma_n^2$ is the noise variance of the measurement and $\sigma^2$ is the variance of the convolved Gaussian, also called the \textit{temperature}. We provide a demonstration of \autoref{eq:likelihood_conv}  in \autoref{sec:likelihood-demo}.

Note that the MCMC algorithm does not navigate the posterior distribution convolved at temperature $\sigma^2$ at the same rate than the likelihood or the prior. Indeed, convolving the product of two distributions is not equivalent to convolving each distribution independently. However, we found empirically that the annealed HMC converges towards the posterior distribution $p(\xb|\yb)$ at zero temperature. We first demonstrate it with a Gaussian posterior in \autoref{sec:gaussian-prior} and then with the full posterior distribution in \autoref{sec:full-prior-desc}, \autoref{sec:post-mean} and \autoref{sec:full-prior-detection}.

This way, the complete sampling procedure consists of running a MCMC on a proxy of the posterior distribution, that is gradually annealed to low temperature, with chains progressively moving toward the posterior distribution. Each chain is initialized by sampling a multivariate normal distribution of unit variance and we leverage the usage of GPUs by running several chains in batches.

\section{Simulations and data} \label{sec:simulations}

In this section we describe the simulated data we used to validate our method, designed to emulate the COSMOS field on which we aim to apply our algorithm.

\subsection{COSMOS shear catalog}
\label{sec:cosmos-data}

\begin{figure}
    \centering
    \includegraphics[width=\linewidth]{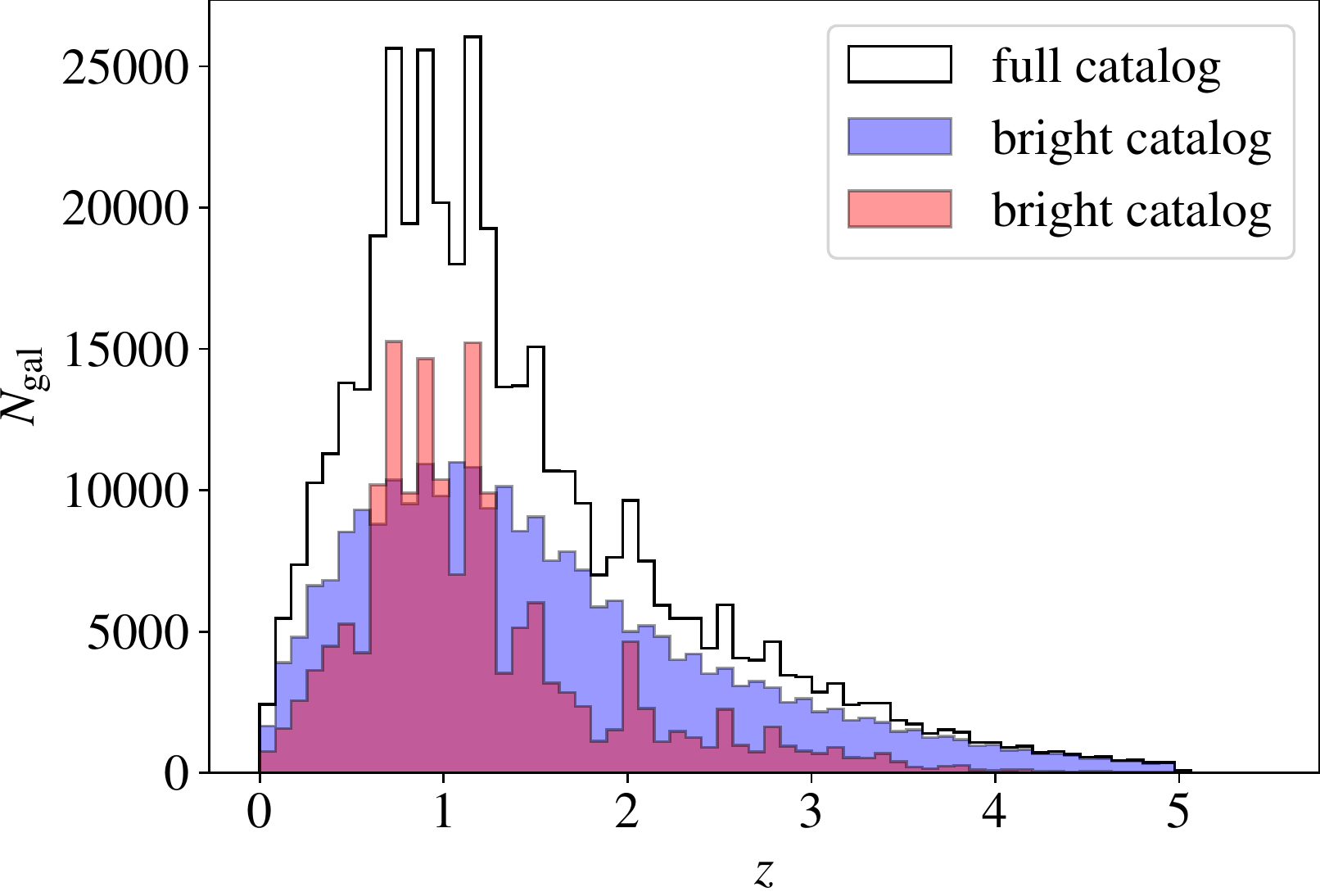}
    \caption{Redshift distribution of the COSMOS catalog from \citet{Schrabback2010}. The redshifts for bright galaxies ($i^{+}<25$) are derived from the COSMOS-30 catalog \citep{Ilbert2009}, while the $n(z)$ distribution for faint galaxies ($i^{+}>25$) is based on an extrapolation to fainter magnitudes as described in \citet{Schrabback2010}. \nblink{COSMOS_shape_catalog}}
    \label{fig:cosmos-redshift}
\end{figure}

The COSMOS Survey \citep{Scoville2007} is a contiguous 1.64 deg$^2$ field imaged with the Advanced Camera for Surveys (ACS) in the F814W band. In this paper, we use the shape catalog obtained in  \citet{Schrabback2010} (hereafter S10) by reduction of this survey using the KSB+ method \citep{Schrabback2007,Erben2001HowShear,Kaiser1995, Hoekstra1998}, applying modelling for the variable HST/ACS point-spread function using principal component analysis and a correction for multiplicative shear measurement bias, which was calibrated as a function of the galaxy signal-to-noise ratio using image simulations.

The S10 catalog is divided into a bright $i^{+}<25$ (Subaru \texttt{SExtractor MAG\_AUTO}
magnitude) and a faint $i^{+} > 25$ galaxy sample. For the bright sample, individual high quality photometric redshifts are available by cross-matching against the COSMOS-30 catalog \citep{Ilbert2009}, while for the faint sample \cite{Schrabback2010} propose a functional form for the overall $n(z)$ based on an extrapolation to fainter magnitudes of the $i_{814}$-redshift relation observed in the $23 <i_{814} <25$ range. The redshift distribution of 
both samples is illustrated on \autoref{fig:cosmos-redshift}.

In this analysis, we combine both bright and faint galaxy samples
into a single shape catalog, and will assume the combined redshift distribution illustrated on \autoref{fig:cosmos-redshift}. The only cut we apply is to reject galaxies in the bright sample with $z_\mathrm{phot}<0.6$ and $i^+>24$. The photometric redshift cut is motivated by S10 finding indications that many of these galaxies are truly at high redshifts. Thus, their inclusion would imply that the used estimate of the redshift distribution is inaccurate. This yields a total of 417117 galaxies, which translates into a mean number of galaxies 64.2 per arcmin$^2$ 

\subsection{$\kappa$TNG simulations} \label{sec:ktng}

 After having described the COSMOS field, we now present the $\kp$TNG suite of simulations we used to create mock data. We use the $\kappa$TNG simulated convergence maps~\citep{Osato2021}, generated from the IllustrisTNG (TNG) hydrodynamic simulations~\citep{Nelson2019,Pillepich2018,Nelson2018,Springel2018,Naiman2018,Marinacci2018}. The TNG simulations are a set of cosmological, large-scale gravity and magneto-hydrodynamical simulations, where baryonic processes such as stellar evolution, chemical enrichment, gas cooling, supernovae and black hole feedback are incorporated as subgrid models. 

The $\kappa$TNG suite is generated through tracing the trajectories of light-rays  from $z=0$ to the target source redshift. A large number of realizations are generated by randomly translating and rotating the snapshots. Specifically, the full set includes 10,000 realisations of $5 \times 5 \, \mathrm{deg}^2$ convergence maps for 40 source redshifts up to $z_s = 2.6$, at a 0.29 arcmin pixel resolution. A flat $\Lambda$-cold dark matter cosmology is assumed for both TNG and $\kappa$TNG: Hubble constant $H_0 = 67.74 \, \mathrm{km} \, \mathrm{s}^{-1} \, {\rm Mpc}^{-1}$,
baryon density $\Omega_\mathrm{b} = 0.0486$,
matter density $\Omega_\mathrm{m} = 0.3089$,
spectral index of scalar perturbations $n_\mathrm{s} = 0.9667$,
and amplitude of matter fluctuations at $8\,\hMpc$ $\sigma_8 = 0.8159$.

\subsection{Mock COSMOS data} \label{sec:emulated-cosmos-data}

To act as our prior for the reconstruction of the COSMOS field, and for validation tests, we emulate from $\kp$TNG simulations mock COSMOS lensing catalogs.

Using the binned galaxy distribution from the S10 catalog at the same 0.29 arcmin resolution, we define a binary mask that captures the limits of the survey as well as missing data within the survey area due to bright stars or image artifacts that prevent the reliable measurement of galaxy shapes in some regions of the survey. This binning also yields noise variance maps, defined by the standard deviation of intrinsic galaxy ellipticities, rescaled by the number of galaxies per pixels $N_i$:
\begin{equation} \label{eq:noise}
\Sigma_{(i,i)} = \left\lbrace \begin{matrix} \frac{\sigma_e^2}{N_{i}} &\mbox{ if $N_{i}>0$}\\
                10^{10} &\mbox{ otherwise}
                \end{matrix} 
                \right.
\end{equation}
for each pixel $i$. Note that in empty pixels, we assume a large variance instead of infinity.

$\kp$TNG provides finely sampled convergence source planes in the range $0<z_s<2.6$. We combine these source planes to match the redshift distribution of the survey illustrated in \autoref{fig:cosmos-redshift}. To handle source redshifts higher than $z_s\geq2.6$ we resorted to a redshift recycling approach in which the last source plane at $z=2.6$ was reused, with an appropriate weight designed to match the expected lensing kernel. This procedure yields a total of 10,000 convergence maps, now properly weighted to match the redshift distribution of the binned data.

Mock observations of COSMOS shear maps are obtained by first computing the shear from simulated convergence maps, sampling spatially variant noise according to $\Sigma$, and applying the binary mask to mask out unobserved regions.

\section{Tests with simulated data} \label{sec:validation}

In this section, we validate our sampling procedure on mock COSMOS data generated as described in the previous section. We will begin by demonstrating our sampling method in an analytically tractable Gaussian case, before testing full posterior sampling using the neural score matching approach introduced in \autoref{sec:nsm}.

\subsection{Sampling validation with analytic Gaussian prior} \label{sec:gaussian-prior}

In this section, we assume that the convergence $\kp$ is a Gaussian Random Field and therefore associated to a Gaussian prior. We show that one can compute the Wiener filter estimate using our posterior sampling method.

Our method, based on the evaluation of the gradient of the log posterior enables to recover both the maximum \textit{a posteriori} (MAP) and the Mean Posterior, which should match.

We used the halofit matter power spectrum from \texttt{jax-cosmo}\footnote{https://github.com/DifferentiableUniverseInitiative/jax\_cosmo}, using \citet{PlanckCollaboration2015PlanckSimulations} (Table 4 final column) results, to build the Gaussian prior covariance matrix. In the following, we also refer to it as the \textit{fiducial} power spectrum.
From the score function of the Gaussian posterior, i.e. $\score (\kp|\g)$, one can compute the MAP using the gradient descent algorithm. As the posterior distribution is Gaussian, there is only one minimum so we are sure to obtain the MAP. The MAP matches the posterior mean, so this also provides a check of the validity of our sampling procedure.

\begin{figure}[h]
    \centering
    \includegraphics[width=\linewidth]{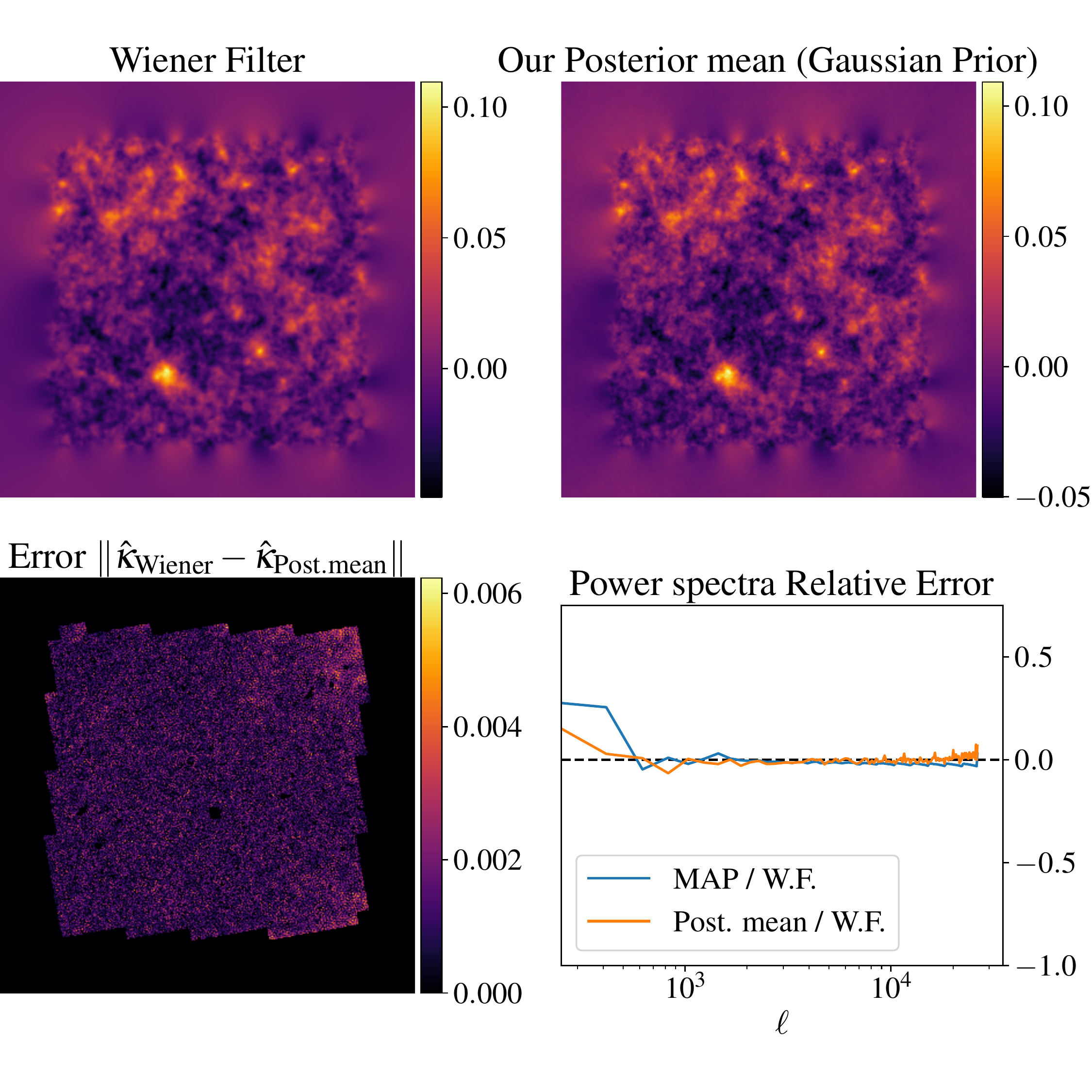}\\
    \includegraphics[width=\linewidth]{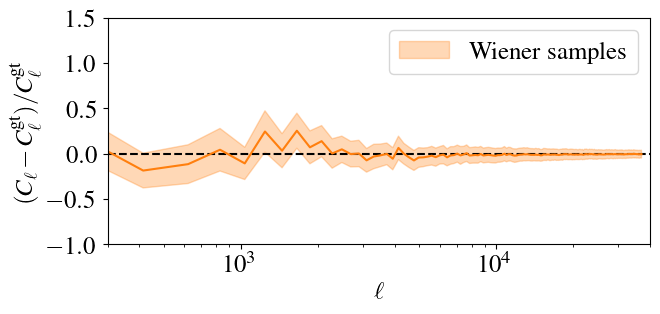} \\
    \caption{Comparison between the Wiener filter (top left) and our score-based maximum-a-posteriori (not shown) and posterior mean reconstructions (top right, average of 1500 samples), assuming a Gaussian prior. The MAP reconstruction matches the Wiener filter both in terms to absolute pixel error and power spectrum ratio (bottom right, blue line). Our posterior mean is also in excellent agreement with the Wiener filter as illustrated by the pixel absolute error (bottom left) and the power spectrum relative error (bottom right, orange line). Bottom panel shows the relative error between the Gaussian posterior samples power spectra and the theoretical power spectrum.
    \nblink{WienerGaussianPrior}}
    \label{fig:gaussian-prior}
\end{figure}

Figure \ref{fig:gaussian-prior} compares the Wiener Filter computed with the messenger field method, from \cite{elsner2013}, and our Score-based MAP and Posterior Mean reconstructions for the same input shear field. The residuals between the MAP and the Wiener filter are very low (two orders of magnitude smaller than the signal) and the residals are also unstructured as expected, both are computed with the same input data and fiducial cosmology and should mathematically match. We also show the relative error between power spectrum of the Wiener filter and the MAP, and the relative error between the Wiener filter and the posterior mean, which is almost zero at every scale. The posterior mean is computed by averaging 1500 posterior samples, sampled with annealed HMC presented in \autoref{sec:sampling-method}, reaching $10^{-3}$ temperature and then projected on the posterior distribution with the ODE sampler.

This experiment with a Gaussian prior validates that our sampling procedure is well adapted to sample from the posterior distribution in this analytically tractable case.

\subsection{Tests with a simulation-based prior}
\subsubsection{Sampling with a neural prior} \label{sec:full-prior-desc}

\begin{figure}
    \centering
    \includegraphics[width=\linewidth]{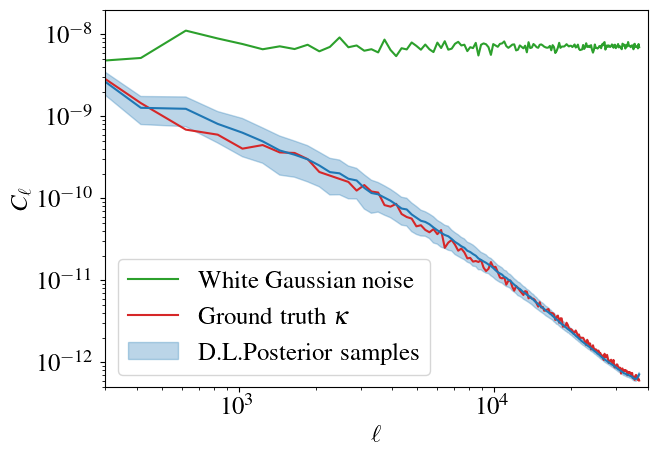} \\

    \includegraphics[width=\linewidth]{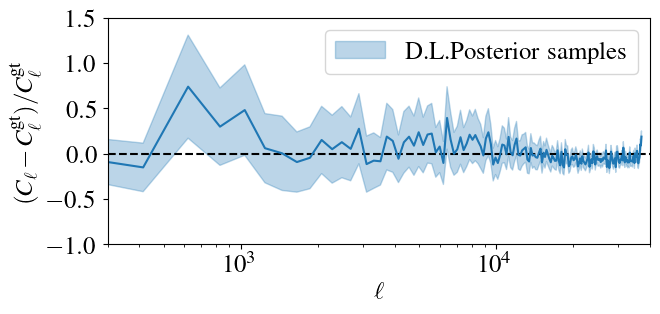} \\
    \caption{Comparison of posterior samples for the Gaussian prior and our hybrid prior. The \textit{dashed black} curve is the fiducial power spectrum, computed using \texttt{jax-cosmo}. The \textit{red} curve is computed from the $\kp$TNG simulated map. The \textit{green} curve is the power spectra of a white Gaussian noise realization we use to initialize the chains. The  \textit{blue} curve is the mean power spectrum of posterior samples assuming our proposed hybrid prior, and the blue interval represent the standard deviation of the power spectra computed over 400 samples. The p.s. of a posterior sample assuming a Gaussian prior is shown in \textit{orange}. Both posteriors consider the same noisy input shear taken from our test set. \nblink{MassMappingResults}
    }
    \label{fig:samples-ps}
\end{figure}

In the last section, we showed that with the gradient of the log-Gaussian posterior can sample the posterior distribution and that we can therefore recover classical results such as the Wiener filter estimate. In this section, we apply the same sampling procedure, using a simulation-based prior.

We saw in \autoref{fig:denoiser} that denoising a convergence map with a deep neural network trained on $\kp$TNG simulated maps (using the \textit{full prior}) is much more effective for small scales structures inference than using the Gaussian assumption alone. We will now show that the same results hold for posterior samples (using the \textit{full posterior}) with a neural prior. The likelihood remains exactly the same as the one used in the former section, i.e. from \autoref{eq:likelihood_conv}. Likewise, the sampling procedure is unchanged, only the prior is extended with a neural network as defined in \autoref{eq:full-prior}.

Convergence maps sampled from the full posterior are shown in the second row of \autoref{fig:methods-maps}. One can observe two regimes in these samples. On one hand, sampling in the unobserved regions (outside of the white boundary of the survey) is only driven by the prior. These regions have therefore high variance between samples because of the different chain initialization maps. On the other hand, in the data region, the sampling is driven by both the likelihood and the prior. Therefore, there is a high correlation between the samples, and with the ground truth convergence map, within the white contours.

\autoref{fig:samples-ps} illustrates that the power spectra of our \texttt{DLPosterior} samples is consistent with the ground truth $\kp$TNG simulated convergence map power spectrum, validating that we are able to sample maps with the correct statistics.
This is also apparent by visual inspection, posterior samples shown in \autoref{fig:methods-maps} are very similar to convergence maps from the $\kappa$TNG simulations, illustrating that we are able to sample $\kp$ maps at the simulation resolution. In addition, \autoref{fig:samples-ps} demonstrates that the prior choice has a strong influence on the map statistics. Using the Gaussian prior alone leads to sampling a map that matches the fiducial power spectrum, which has more power at high $\ell$ due to the limited resolution of the simulation.

\subsubsection{Posterior mean} \label{sec:post-mean}

\begin{figure*}[h!]
    \centering
    
    \includegraphics[width=\textwidth]{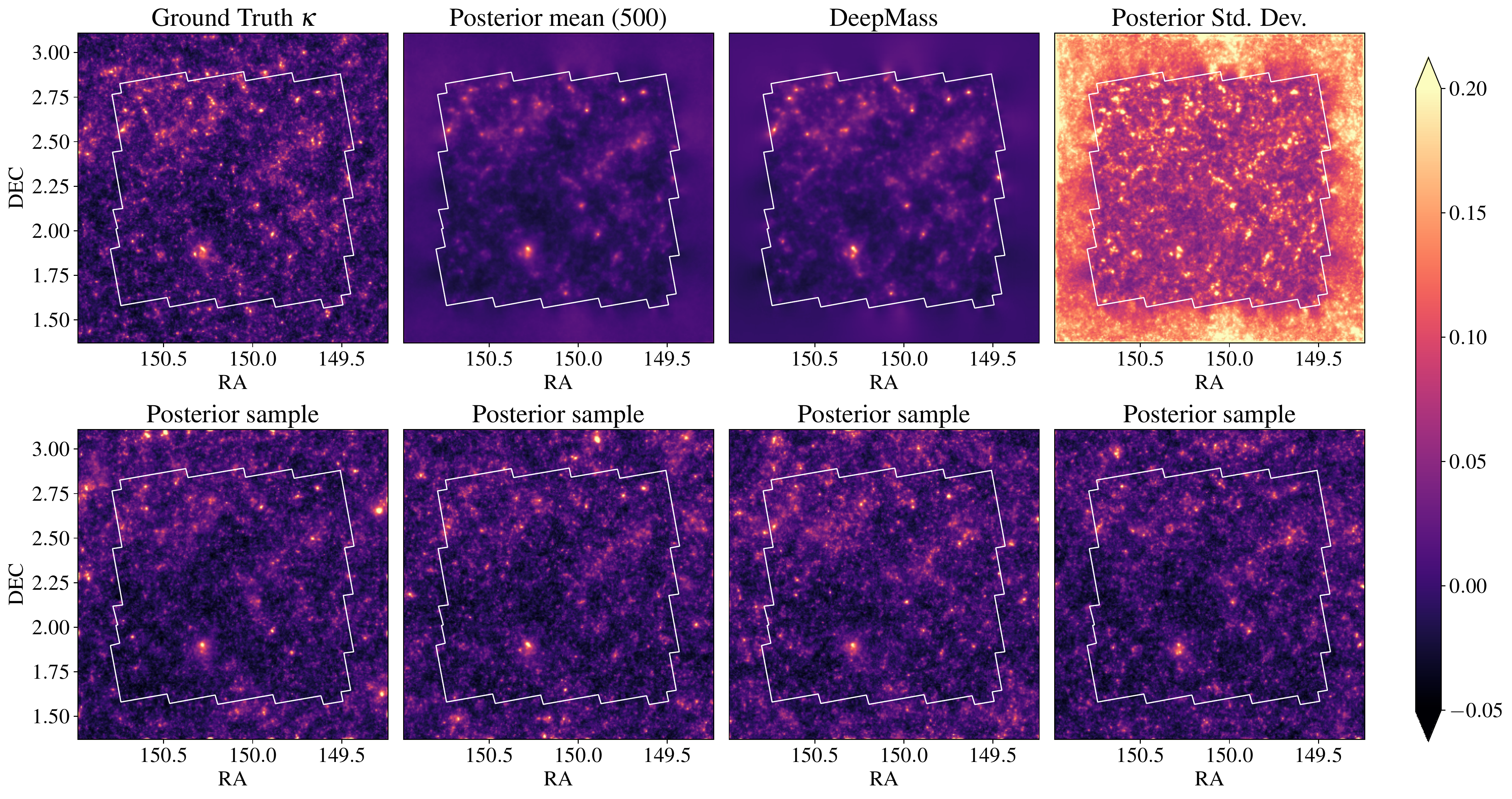}
    \caption{Comparison of deep learning-based mass-mapping methods on one mocked COSMOS field. \textit{First row}: the ground truth $\kp$TNG convergence map, the mean of our posterior samples (over 400 samples), DeepMass and the standard deviation of our posterior samples. All the maps use the same colorbar, except for the standard deviation, which is displayed in the range $[0,0.035]$. \textit{Second row}: samples from the posterior distribution, using the hybrid Gaussian-neural score prior. The input noise level is computed according \autoref{eq:noise} and the mask contours corresponds to the COSMOS survey boundary mask. \nblink{MassMappingResults}}
    \label{fig:methods-maps}
\end{figure*}

As for the Wiener filter, a way to validate the posterior distribution is to look at the posterior mean solution. In this section, we compare to the state of the art method to date, \texttt{DeepMass}, which is an other deep-learning based method. We trained \texttt{DeepMass} on examples drawn from our prior and likelihood model with the exact same U-net architecture as our denoiser. Using the convergence maps from our simulated dataset, we mocked noisy shear maps according to the likelihood in \autoref{eq:likelihood_conv}. \texttt{DeepMass} computes the posterior mean solution (\autoref{deepmass}), which can be recovered by averaging the posterior samples from our procedure, i.e. $\int \kp p(\kp|\g) d\kp \approx \sum_{\kp\sim p(\kp|\g)} \kp$. Even if \texttt{DeepMass} is learning both the prior and the likelihood, the two posterior means should match since both methods were learned on the same simulations.

A qualitative comparison of the method is discussed in figure \ref{fig:methods-maps}. Table \ref{tab:methods-metrics} shows a quantitative comparison of the Kaiser-Squires, the Wiener Filter, \texttt{MCALens}, \texttt{GLIMPSE}, \texttt{DeepMass} and our mass-mapping method, based on two metrics:
\begin{itemize}
    \item{The root mean square error (RMSE) defined as:

\begin{equation} \label{eq:rmse}
    \text{RMSE}\left(\hat{\kp}, \kp^{gt}\right) = \sqrt{\frac{1}{n}\sum_{i=1}^{n} \left( \hat{\kp}_i - \kp^{gt}_i \right)^2},
\end{equation}

where $i$ is the pixel index, $\hat{\kp}$ the mean-subtracted estimated convergence map, $\kp^{gt}$ the mean-subtracted ground truth convergence map we aim to recover. Note that we use here the mean-subtracted convergence map. This is motivated by the mass-sheet degeneracy, which does not constrain the mean of the convergence map.}

\medskip

\item{The Pearson correlation coefficient $r$ defined as:

\begin{equation} \label{eq:pc}
    r\left(\hat{\kp}, \kp^{gt}\right) = \frac{\text{Cov}\left(\hat{\kp}, \kp^{gt}\right)}{\sigma_{\hat{\kp}} \sigma_{\kp^{gt}}},
\end{equation}

where $\text{Cov}$ is the covariance and $\sigma_\kp$ is the standard deviation of the convergence map $\kp$.
}

\end{itemize}

On the top row of figure \autoref{fig:methods-maps} we display our Posterior mean against \texttt{DeepMass}, the ground truth and the Wiener filter. Qualitatively, we can clearly observe that both \texttt{DeepMass} and our posterior mean recover the convergence map with striking visual similarity, and at a much higher resolution than the Wiener Filter applied to the same field (see \autoref{fig:gaussian-prior}). We can indeed identify the presence of clusters, which are corresponding to true clusters in the ground-truth map. For further visual comparison, \autoref{fig:method-comparison} shows Kaiser-Squires, Wiener Filter, \texttt{MCALens}, \texttt{GLIMPSE}, \texttt{DeepMass}, and our Posterior mean and samples convergence estimates from the same input shear field. 

Just like in the case of Gaussian prior explored in the previous section, one can observe that computing the mean of the posterior samples cancels out the signal outside the survey mask, where the reconstruction is only constrained by the prior. This is expected since sampling with a random initialization yield a random sample on the posterior. This is moreover confirmed by the standard deviation map, showing that the uncertainty of the reconstruction is much higher in masked regions than regions constrained by data.

Quantitatively, we also compare our reconstruction to methods based on other priors, which are \texttt{MCAlens} and \texttt{GLIMPSE}. These methods are based on sparse priors, which are described in \autoref{sparse-prior}. It turns out that \texttt{MCAlens}, \texttt{DeepMass} and our posterior mean both reach the lowest RMSEs and highest Pearson correlations. This results can be expected since both \texttt{DeepMass} and \texttt{DLPosterior} mean are data-driven. \texttt{DeepMass} is explicitly trained to minimize the pixel reconstruction error leading to an estimate of the posterior mean, which our method is also targeting. This result shows that our method reaches the state of the art reconstruction of convergence maps.

The fact that \texttt{MCAlens} leads to metrics similar to \texttt{DeepMass} and \texttt{DLPosterior} mean could be explained by the fact that \texttt{MCAlens} modeling, (the convergence map being modeled as the sum of a Gaussian Random Field and a non-Gaussian one being sparse in the wavelet domain) is able to capture well enough the main properties of the convergence map. 

\medskip

\begin{table}[]
    \centering
        
        \begin{tabular}{lll}
        \hline
            Method & RMSE $\downarrow$ & $r$ $\uparrow$ \\
        \hline
        \hline
            KS \tiny{($\sigma_\text{smooth}$=1 arcmin)} & $2.40\times10^{-2}$ & 0.57 \\ 
            Wiener filter & $2.31\times10^{-2}$ & 0.61\\
            \texttt{GLIMPSE} (3) & $2.84\times10^{-2}$ & 0.42\\
            
            \texttt{MCAlens} (5)& $2.19\times10^{-2}$ & 0.67\\
            \texttt{DeepMass} & $\bm{2.18\times10^{-2}}$ & $\bm{0.68}$\\  
            \texttt{DLPosterior} mean & $\bm{2.16\times10^{-2}}$ & $\bm{0.68}$\\ 
        \hline
        \end{tabular}
    \caption{Metrics comparison between different mass mapping methods. RMSE (the lower the better) is computed according to eq. \ref{eq:rmse} and the Pearson Correlation coefficient $r$ (the higher the better) according to eq. \ref{eq:pc}. For a fair comparison to Kaiser-Squires, we kept the smoothing coefficient that minimized the RMSE.}
    \label{tab:methods-metrics}
\end{table}


\subsection{Demonstration with cluster detection} \label{sec:full-prior-detection}

\begin{figure}
    \centering
    \includegraphics[width=\linewidth]{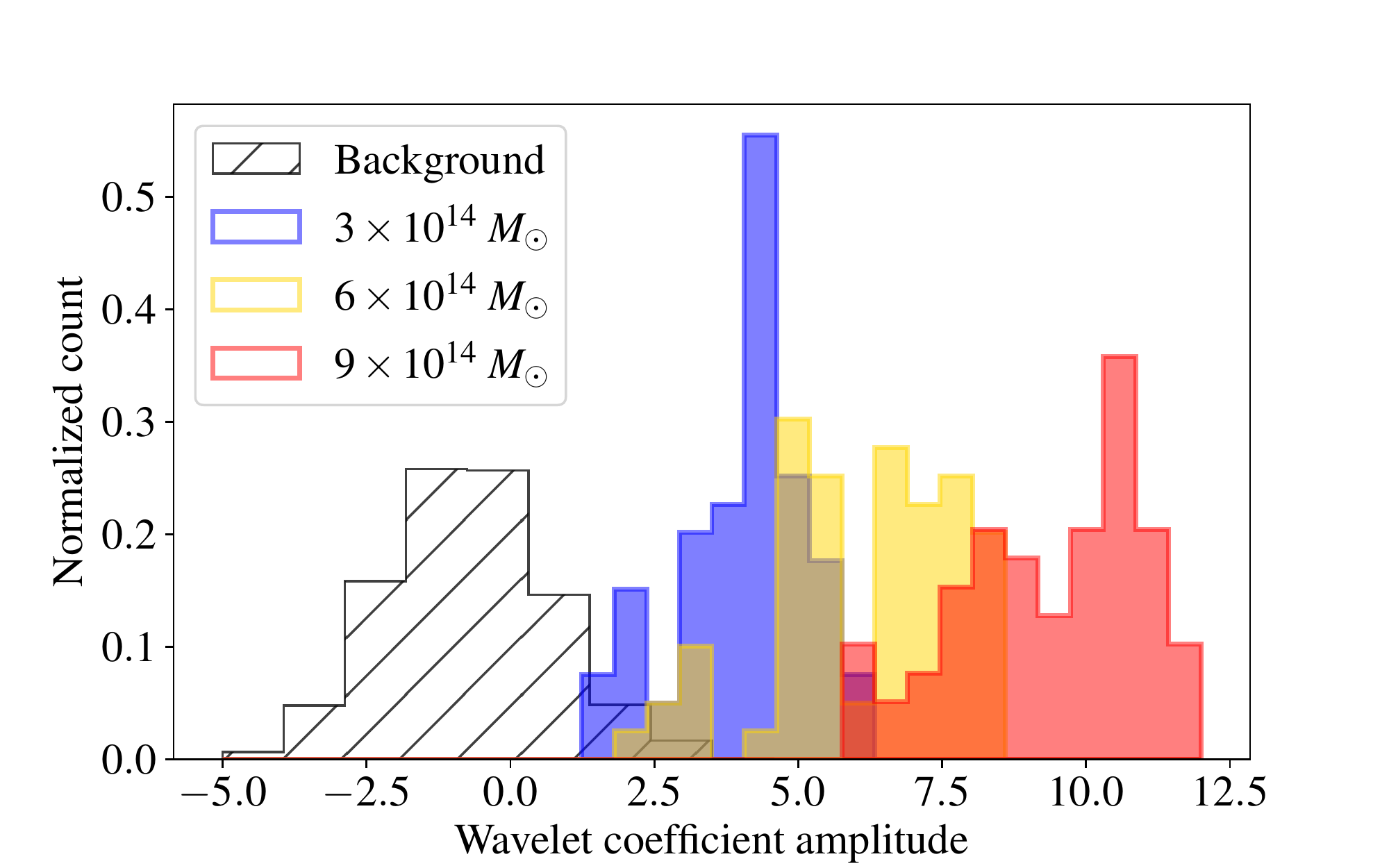}
    \caption{Histograms of coefficient amplitudes. A wavelet coefficient amplitude corresponds to the coefficient computed in \autoref{eq:detection-crit}. Background corresponds to the coefficient of samples where we did not add any cluster. The background coefficients follow a Gaussian distribution, we can thus define a standard deviation $\sigma$ for detection. We can see that the lower the cluster mass \textendash thus the lower the SNR, the closer histograms are to the background. \nblink{Detection}}
    \label{fig:hist_detection}
\end{figure}

\begin{figure}
    \centering
    \includegraphics[width=\linewidth]{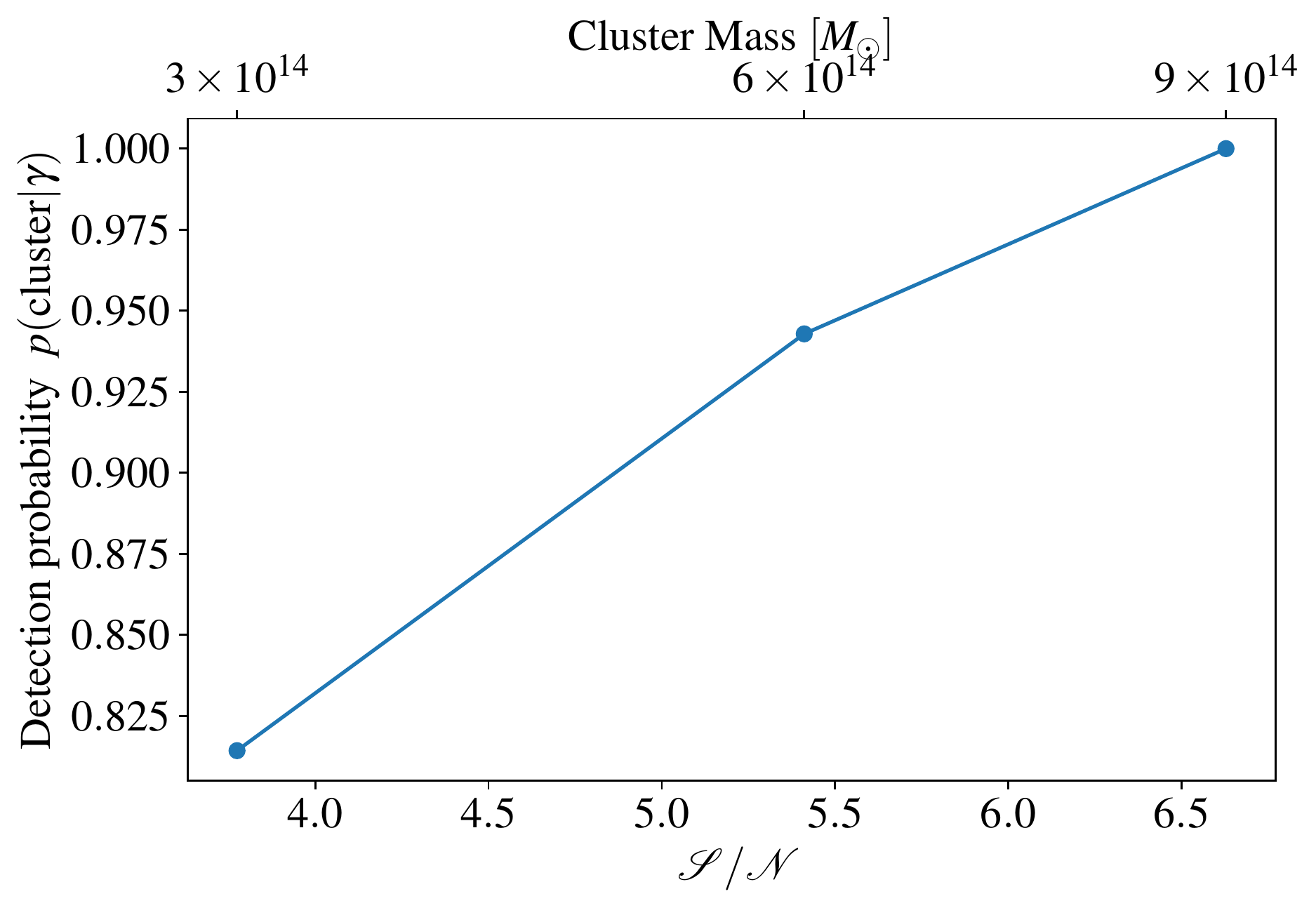}
    \caption{Comparison of the probability of cluster detection given the input data and the detection method with respect to the SNR. The ratio of detection corresponds the the number of samples above 5$\sigma$ over the total number of samples. The SNR was computed by dividing the maximum $\kp$ value of the NFW profile by the noise in the $\kp$ field. We considered clusters at redshift $z_\mathrm{h}=0.5$ and concentration $c=1$ varying the halo mass. \nblink{WienerGaussianPrior}}
    \label{fig:SNR_detection}
\end{figure}

In a Bayesian perspective, we expect that there is a relation between the number of times a cluster appears within posterior samples and its signal-to-noise ratio (SNR). We show in this section to which extent having access to posterior samples can allow us to quantify the uncertainty of a cluster detection and how it relates to the cluster SNR.

In order to compare the occurrence of a cluster in the posterior and its SNR, we insert a Navarro-Frenk-White (NFW) lensing signal, described in \cite{Navarro1997} and  \cite{Takada2003}, of a given mass, redshift, and concentration, into a mocked shear field. We consider a cluster at redshift $z_\mathrm{h}=0.5$, concentration $c=1$, lensing a source at redshift $\mathrm{s}=1$. We run the experiments at different halo masses $\{9\times 10^{13},3\times 10^{14}, 6\times 10^{14}\} ~\solarmass$ that simulates an SNR increase for a given noise level in the data. We used the code from the \texttt{lenspack}\footnote{https://github.com/CosmoStat/lenspack} repository to build the NFW shear map, that is summed to the mocked shear field from our test dataset.

In this section, all the maps are filtered with an aperture mass filter with starlets functions \citep{leonard2012}. Starlets are well localized in real space, hence it is  well very adapted to cluster detection \citep{Starck1998}. In the following, we either refer as pixel or coefficient for the filtered convergence maps. 

Therefore, for a given NFW cluster, its intrinsic SNR, that we also call the input SNR, is defined as: 
\begin{equation}
    \mathrm{SNR}_\text{input} = \frac{\underset{{i,j}}{\max}\,\, \breve \kp_\mathrm{NFW}\left[i,j\right]}{ \mathrm{Std} \left( \breve \kp_\mathrm{background}\right)},
\end{equation}
where $\breve \kp_\mathrm{NFW}$ denotes the convergence NFW profile that has been filtered and the maximum is taken over the pixels of the map, $\breve \kp_\mathrm{background}$ is the background convergence map, i.e. the mocked convergence field over which we add the NFW profile, that is being filtered. The mocked convergence field is computed from COSMOS-like data, similar to the fields described in \autoref{sec:emulated-cosmos-data}.

We now describe the detection procedure. Given a shear field, we sample convergence maps with our method and each posterior sample is filtered with the aperture mass. Given one posterior sample, the detection criteria defined as:

\begin{equation}
    \mathrm{Detection} =\begin{cases}\mathrm{True\,\, if\,}& \frac{\underset{{i,j}}{\max}\,\, \breve \kp_\mathrm{post.\ sample}\left[i,j\right]}{\mathrm{Std} \left( \breve \kp_\mathrm{background}\right)} > 3, \\ \mathrm{False\,}&\mathrm{otherwise},\end{cases}
    \label{eq:detection-crit}
\end{equation}
which means that we consider that a cluster is detected when the convergence is over 3 times the noise background amplitude. The distribution of the convergence background is illustrated as the hashed histogram in \autoref{fig:hist_detection}.  Similar histograms were computed from posterior samples with an added cluster in the input shear and plotted besides the background. In Figure \autoref{fig:hist_detection}, we can see that when decreasing the input SNR, by reducing the halo mass, the distribution of the cluster maximum coefficient shifts towards lower levels, approaching the background distribution. The background distribution histogram in \textit{hatched black} in \autoref{fig:hist_detection} is the histogram of the $\breve \kp_\mathrm{background}$ map. Figure \ref{fig:stamps_detection} shows stamps coefficients after filtering.

Figure \ref{fig:SNR_detection} shows the frequency of the clusters detection in our posterior samples as a function of the input SNR.

\medskip

\section{Reconstruction of the COSMOS field} \label{sec:cosmos-recon}

\begin{figure*}[h!]
    \centering
    \includegraphics[width=\textwidth]{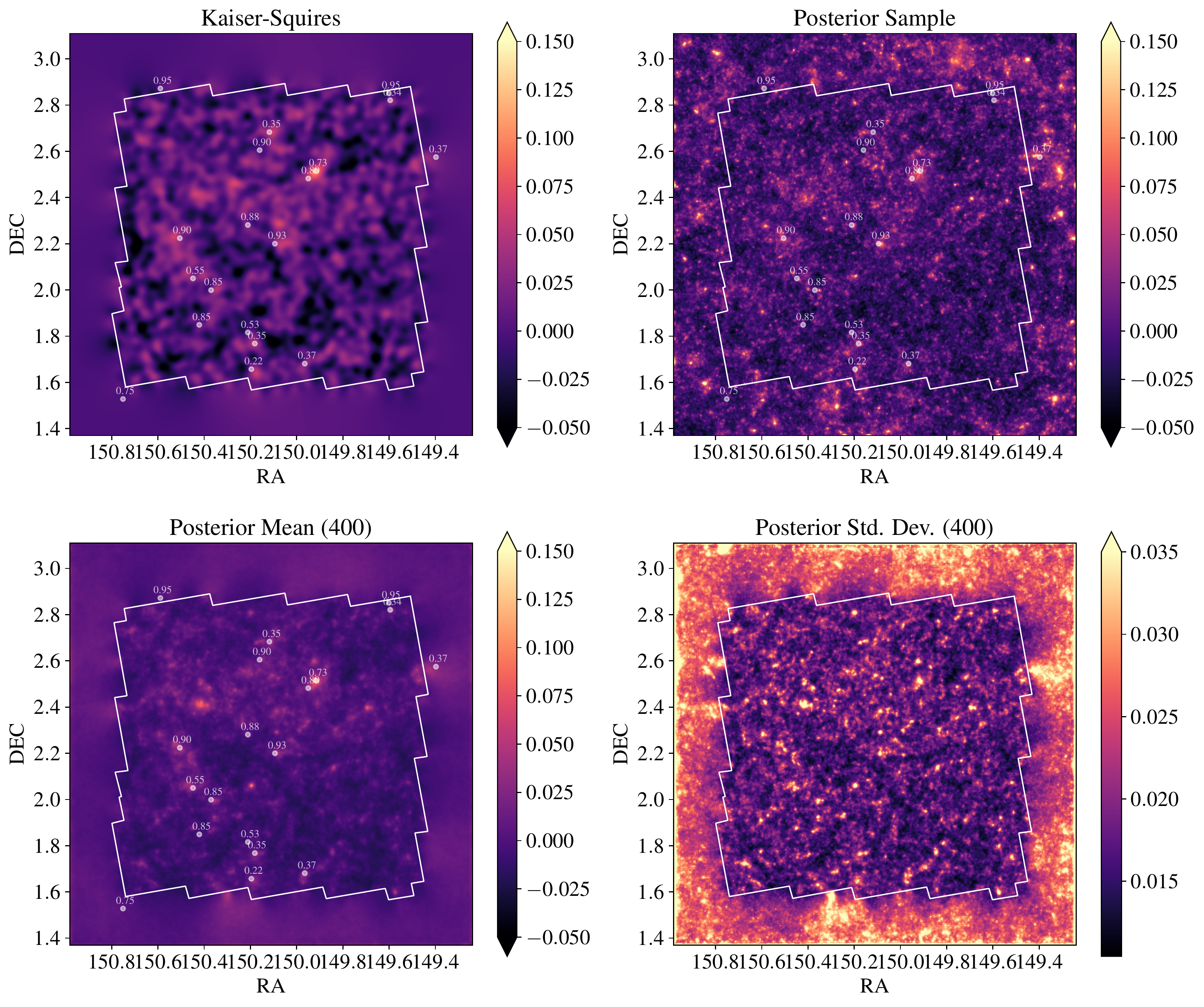}
    \caption{HST/ACS COSMOS field reconstructions along known X-ray clusters from the XMM-Newton survey. \textit{Top-left}: Kaiser-Squires (with a Gaussian smoothing of $\sigma=1$ arcmin), \textit{top-right}: sample from the posterior distribution, \textit{bottom-left}: mean of the posterior distribution, \textit{bottom-right}: standard deviation of the posterior distribution (over 400 samples, shown in the clipped range [0, 0.035]). \nblink{MassMappingResults}}
    \label{fig:cosmos-maps}
\end{figure*}

In this section we now apply our full methodology to the reconstruction of the COSMOS field, using the catalog described in \autoref{sec:cosmos-data}. The likelihood convariance and the simulation-based prior remain the same as in the validation section.

As a baseline, we show the Kaiser-Squires reconstruction of the COSMOS convergence field in the \textit{top-left} panel of \autoref{fig:cosmos-maps}. We applied a Gaussian-smoothing with a variance $\sigma_\text{smooth} = 1$ arcmin, chosen such that the RMSE is minimized and Pearson correlation coefficient is maximized on simulated data. Although large scale and small scale structures can already be observed on this map, its power spectrum does not correspond to the fiducial matter power spectrum, and no feature at scales smaller that 1 arcmin can be observed. Moreover, there is not any uncertainty quantification on the reconstruction.

\begin{figure}[h!]
    \centering
    \includegraphics[width=\linewidth]{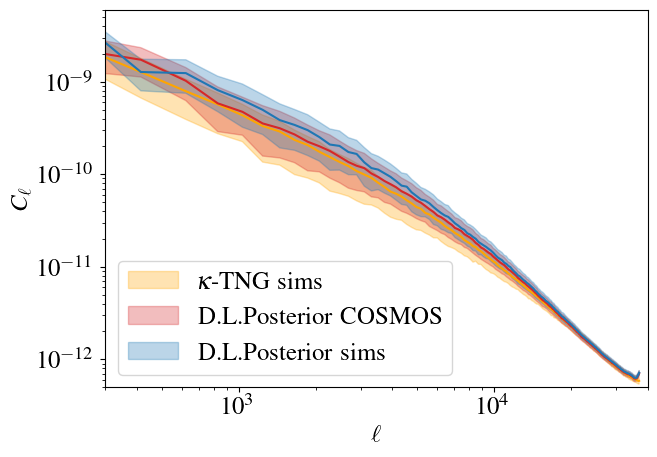}
    \caption{Power spectra comparison between the COSMOS field reconstruction and simulations. We compare power spectra of posterior samples for the COSMOS field in \textit{blue} and for a simulated field in \textit{red}. We also display the power spectrum of samples from the training set in \textit{orange} to show the matching.  \nblink{MassMappingResults}}
    \label{fig:cosmos-ps}
\end{figure}

In \autoref{fig:cosmos-maps} we also presents our reconstruction of the COSMOS convergence field, alongside uncertainty quantification. The \textit{top-right} panel shows a posterior sample that is looking very similar to a posterior sample from simulated input  from $\kp$TNG simulations. Notice that although there is only input shear within the white contours, i.e. in the survey footprint, a complete convergence map is sampled from the posterior distribution. We furthermore validate, in \autoref{fig:cosmos-ps}, the quality of COSMOS posterior samples reconstruction by showing that their power spectra are in good agreement with those of $\kp$TNG convergence maps.

As proposed in \autoref{deepmass} we choose the posterior mean as our estimate for the convergence reconstruction. In the \textit{Bottom-left} panel of \autoref{fig:cosmos-maps} shows the average of 400 samples making the \texttt{DLPosterior} mean of the COSMOS field. One can visually observe the similar large scale structures of the field between the Kaiser-Squires and the \texttt{DLPosterior} mean, while the latter contains much more resolved features, especially concerning the clusters shapes.

Another way to examine our COSMOS convergence map is to compare to another probe for cluster mass detection. So in the \textit{bottom-left} and \textit{top-right} panels, we overlay a subset of X-ray clusters from the \citet{Finoguenov2006} catalog. Most of the X-ray clusters match to a resolved peak in the convergence field, which is in a way expected since we selected the most massive X-ray clusters.

Alongside the \texttt{DLPosterior} mean, we also provide the \texttt{DLPosterior} standard deviation, in the \textit{bottom-right} panel of \autoref{fig:cosmos-maps}. One can observe again that the variance is the highest outside the survey contours since there is no data. Another interesting behaviour of the posterior is that the location where the uncertainty is also high is where the signal is the most intense.

The former mass-map of the COSMOS field was published in \citet{Massey2007DarkScaffolding}, using an generalized version of the Kaiser-Squires method. When comparing the two maps, although we do not share the same shape catalog nor redshift distribution, they clearly share similar mass distributions. However, the COSMOS-\texttt{DLPosterior} mean is much more resolved. In particular, we can identify several clusters in what looks like one coarse cluster in the \citet{Massey2007DarkScaffolding} mass-map at coordinates RA=150.4, DEC=2.5.

\section{Discussion}
Having presented the method and results, we propose in this section a discussion on some important points including scientifically relevant use-cases, limitations, and possible extensions.

As mentioned earlier in this paper, taking the average over posterior samples should reduce to the \texttt{DeepMass} results. One may wonder about the tradeoff between the two approaches if the results should be the same. We would argue that the two methods are complementary. \texttt{DeepMass} is much faster in producing a convergence map as it only requires a forward pass of the U-net. However, \texttt{DeepMass} remains an \textit{amortized} solution, with no strong guarantees on the solution it recovers in practice, due to not having an explicit likelihood. This also means that the entire model needs to be retrained for any change in the lensing catalog (to account for variations in the mask or noise).

We expect that the method presented in this paper will find its most compelling applications in the study of localized structures through the weak lensing effect, where the benefits of a full pixel-level posterior are the strongest. As a particularly relevant example, we will mention the discussion surrounding the potential detection in the Abell 520 (A520) cluster of a dark clump  \citep{Jee2012, Clowe2012}, a localized peak visible in mass-maps but with no optical counterpart. Quantifying the significativity of such a structure using non-linear mass-mapping algorithms targeting a Maximum A Posterior solution (for instance based on sparse regularisation) is a difficult task. It was attempted for this particular field using different techniques in \citet{Peel2017} and \citet{Price2018}, but without strong quantitative statements. The method developed in this paper however would be able to access the full posterior of the problem.

For cosmological applications, it is however likely that the use cases of the method presented in this work will remain limited. Indeed, for Higher Order Statistics relying on simulations to evaluate their likelihood, the particular mass-mapping technique used is not critical as any systematics due to reconstruction errors induced by the algorithm are calibrated on simulations. A simple Kaiser-Squires inversion should in principle suffice in most cases of interest. From an information theoretic point of view, the information is preserved by a Kaiser-Squires inversion (in the presence of masks, keeping both E- and B- modes) while any posterior summary may discard some cosmological information. 
Nevertheless, the ability to sample constrained realisations may find very useful applications such as inpainting masked regions for the purpose of facilitating the computation of 3pt functions, or void detection algorithms.

We also want to highlight that the method presented in this paper can be extended in multiple directions. We are considering here a prior at a fixed cosmology, primarily due to the high computational cost of high resolution hydrodynamical simulations at different points in cosmological parameter space. Concretely, this means that our solution is heavily biased towards the fiducial cosmology on scales poorly constrained by the data. Note however that the hybrid deep learning + physical gaussian prior we have introduced in this work provides a natural framework for extending our method to include cosmological dependence. On large scales, the prior is mostly driven by the analytic power spectrum and thus easy on condition on cosmology, while on small scales, only the non-Gaussian residuals are captured by the neural network. This implies that the cosmology-dependent part  of the model that needs to be learned on simulations is mostly on small scales. This makes it very likely that in the close future one could develop such a cosmology-dependent residual prior from suites of numerical simulations of smaller cosmological volume, but spanning a range of cosmological models. The recent multifield CAMELS simulations \citep{Villaescusa2021} would be an ideal dataset for this purpose. Another avenue for future work would be extending the method to the sphere, which is simply a matter of defining a U-net for instance using a DeepSphere \citep{Perraudin2019} approach for convolutions on a spherical domain.

\section{Conclusion}

In this paper, we have presented a unified view of the mass-mapping problem as a Bayesian posterior estimation problem. Most existing methods either rely on simple priors (i.e. Gaussian prior for the Wiener filter) and/or only return a point estimate of the mass-mapping posterior. Instead, we have proposed a framework which allows us to 1) use numerical simulations to provide a full non-Gaussian prior on the convergence field; 2) sample from the full Bayesian posterior of the mass-mapping problem under this simulation-driven prior and a physical likelihood.

The proposed approach, dubbed \texttt{DLPosterior}, relies first on using a Denoising Score Matching technique to learn a prior from high resolution hydrodynamical simulations (the $\kappa$TNG dataset \citep{Osato2021}), an approach that has proven to be extremely scalable and easy to implement. And as a second ingredient, we have introduce an annealed Hamiltonian Monte-Carlo approach that allows us to sample the high-dimensional Bayesian posterior of the problem with high efficiency. We demonstrate that we are able to achieve on average an independent posterior sample in 10 GPU-minute on an Nvidia Tesla V100 GPU. Moreover, the annealing scheme ensures that we sample independent samples with the correct weights of the posterior modes.

We first validated the sampling approach in an analytically tractable fully Gaussian case where we recovered the Wiener filter. We then validated the entire method on mock observations, by comparing the posterior mean obtained by our method to a Deep Learning estimate of this posterior mean using the \texttt{DeepMass} method. We found excellent agreement between these two independently estimated posterior summaries, with near identical Pearson correlation coefficient and RMSE. This consistency gives us high confidence that the full posterior is correctly sampled even in the non-Gaussian case.

In further comparisons, \texttt{DLPosterior}, demonstrated quantitative improvement on convergence reconstructions against other standard methods, based on a large class of priors such as the Kaiser-Squires inversion, the Wiener Filter, or \texttt{GLIMPSE2D}. In addition, contrary to most of these methods, \texttt{DLPosterior} provides uncertainty quantification with the posterior samples, such as the posterior variance. In this respect, we also showed that the recovered posterior can be interpreted to quantify uncertainties, by establishing a close correlation between posterior convergence values and the SNR for clusters artificially introduced into a field.

Finally, we have applied the validated method to the reconstruction of the HST/ACS COSMOS field based on the shape catalog from \citet{Schrabback2010}, and produced the highest quality convergence map of this field to date.

\bigskip

In the spirit of reproducible and reusable research, all software, scripts, and analysis notebooks are publicly available at:\\

\url{https://github.com/CosmoStat/jax-lensing}

\bigskip

\textit{\small Software: Astropy \citep{Robitaille2013, PriceWhelan2018}, IPython \citep{Perez2007}, Jupyter \citep{Kluyver2016}, Matplotlib \citep{Hunter2007}, Numpy \citep{2020NumPy-Array}, TensorFlow Probability \citep{Dillon2017}, JAX \citep{jax} }

\begin{acknowledgements}
The authors are grateful Zaccharie Ramzi for fruitful discussions on inverse problems and very helpful discussions on neural network architecture.
Benjamin Remy acknowledges support by the Centre National d’Etudes Spatiales and the Universit\'e Paris-Saclay Doctoral Program
in Artificial Intelligence (UDOPIA).
KO is supported by JSPS Research Fellowships for Young Scientists.
This work was supported by Grant-in-Aid for JSPS Fellows Grant Number JP21J00011.
This work used the Extreme Science and Engineering Discovery Environment (XSEDE),
which is supported by NSF Grant No.~ACI-1053575. This work was granted access to the HPC resources of IDRIS under the allocation 2021-AD011011554R1 made by GENCI (Grand Equipement National de Calcul Intensif). We thank New Mexico State University (USA) and Instituto de Astrofisica de Andalucia CSIC (Spain) for hosting the Skies \& Universes site for cosmological simulation products.

\end{acknowledgements}

\newpage

\bibliographystyle{aa-files/bibtex/aa}
\bibliography{references,references2,JLSBibTex}

\begin{appendix}

\section{Score-based Metropolis-Hastings} \label{sec:MH}

The discretized implementation of the Hamiltonian Mote Carlo algorithm requires to correct for the discretization error. The chain update computed with \autoref{eq:hmc} is a proposal that is accepted with probability of the form $\min \{ 1, p(x_*)q(x_n|x_*) / p(x_n)q(x_*|x_n) \}$, where $x_n$ is the last sample of the chain, $x_*$ is the HMC proposal, $p$ is the target density and $q$ is a proposal distribution from a random walk, i.e. $q(x_n|x_*) = \mathcal{N}(x_n|x_*, \boldsymbol{\mathrm{M}})$ with $\boldsymbol{\mathrm{M}}$ the HMC preconditionig matrix.

In our approach, we do not directly have access to the distribution $p$, but to its score function $\score$. However we can still approximate the log ratio needed to compute the MH acceptance probability from the scores using the path integral:
 
\begin{equation} \label{eq:score-MH}
    \log p(x_*) - \log p(x_n) = \int_0^1 \nabla_x \log p (t*(x_*-x_n) + x_n) \cdot (x_* - x_n) dt,
\end{equation}

which we evaluate in practice with a simple 4 points Simpson integration rule. This integral could be approximated to any precision at the cost of additional score evaluations.

Thus, with \autoref{eq:score-MH} at hand, we are able to implement a large class of MCMC algorithm such as HMC or Metopolis Adjusted Langevin Algorithm, using the score function only.

\section{Sampling by solving a Stochastic Differential Equation} \label{sec:ode}

Recent work from \cite{Song2020a} improves annealed sampling procedures by generalizing the process as a Stochastic Differential Equation (SDE). A SDE has the following form:
\begin{equation}\label{eq:SDE}
    d\xb = f(\xb, t)dt + g(t)d\wb,
\end{equation}
where $f(\cdot,t):\mathbb{R}^d\rightarrow\mathbb{R}$ is a vector valued function, $g(\cdot):\mathbb{R}\rightarrow\mathbb{R}$ a scalar function, and $\wb$ a Wiener process. This equation models a diffusion process $\{\xb(t)\}_{t=0}^T$ indexed by a continuous time variable $t\in[0,T]$, such that $\xb(0)\sim p_0$ and $\xb(T)\sim p_T$, where $p_0$ denotes the data distribution and $p_T$ the convolution between the data distribution and a multivariate Gaussian of variance $T$. The variance $T$ being much larger than the support size of the data distribution, $p_T$ can be seen as a wide multivariate Gaussian.

A result from \cite{Anderson-RSDE} shows that one can reverse the SDE in \autoref{eq:SDE}, starting from samples $\xb(T)\sim p_T$ and obtaining samples $\xb(0)\sim p_0$, by computing the reverse-time SDE, involving the \textit{score function}:
\begin{equation}\label{eq:reverse-SDE}
    d\xb = \left[f(\xb, t) - g(t)^2 \nabla_{\xb} \log p_t(\xb)\right]dt + g(t)d\bar{\wb},
\end{equation}
where $\bar{\wb}$ is a Wiener process with backward time, from $T$ to $0$.

\cite{Song2020a} stress that the annealed LD is a discretized version of a SDE. Indeed, the chain update of the annealed LD is:
\begin{equation}
    \xb_i = \xb_{i-1} + \sqrt{\sigma_i^2-\sigma_{i-1}^2} \zb_{i-1},
\end{equation}
with $i \in \{1,\dots,N\}$, where the $\{\sigma_i\}_{i=1}^N$ are the increasing variance of the Gaussian we convolve the data distribution, also called the temperature and $z_i$ are multivariate Gaussian realization. Then if $N\rightarrow\infty$, the temperature becomes a continuous function $\sigma(t)$ and $\zb_i$ becomes a Gaussian process $\wb(t)$. Thus, the process $\{\xb(t)\}_{t=0}^T$ is given by the SDE:
\begin{equation}
    d\xb = \sqrt{\frac{d}{dt}\Big(\sigma^2(t)\Big)} d\wb.
\end{equation}

Finally, \cite{Song2020a} demonstrate that to any reverse-time SDE expressed as in \autoref{eq:reverse-SDE}, corresponds a deterministic process whose trajectory has the same marginal probability densities $\{p_t(\xb)\}_{t=0}^T$, satisfying an ODE:
\begin{equation}\label{eq:ODE}
    d\xb = \left[f(\xb, t) - \frac{1}{2}g(t)^2 \nabla_{\xb}\log p_t(\xb)\right]dt,
\end{equation}
with the exact same notation as in the SDE equation. Thus, in order to sample from the distribution $p_0$, one can solve the corresponding ODE of the annealed LD:
\begin{equation}
    d\xb = -\frac{1}{2}\sqrt{\frac{d}{dt}\Big(\sigma^2(t)\Big)}\nabla_{\xb}\log p_t(\xb)dt,
\end{equation}
using any black-box ODE solver. Notice that, once again, the sampling procedure only depends on the score function. Thus, one can to sample from the target distribution, denoted $p_0$ in this formalism, starting from random samples from a multivariate Gaussian $p_T$.

\section{U-net architecture}
\label{sec:unet}

We used a 3-scale U-net architecture \citep{Ronneberger}, composed of residual blocks \citep{He2016a}. Each block is composed of 2 convolutions followed by a Batch Normalization and 3 convolutions for the bottleneck. Each batch normalisation is followed by a ReLU non-linearity except the last one. As densigned in \cite{Ronneberger}, we performed downsampling by average pooling and upsampling by upconvolutions. We used the following sequence of channels $[32,64,128,128]$ over the different scales.

\begin{figure}[h!]
    \centering
\begin{tabular}{cc}
    \includegraphics[width=0.45\linewidth]{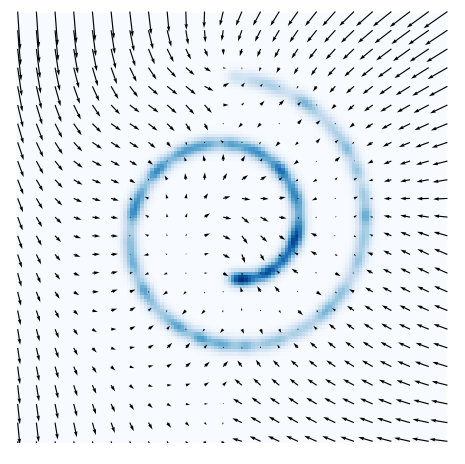} & \includegraphics[width=0.45\linewidth]{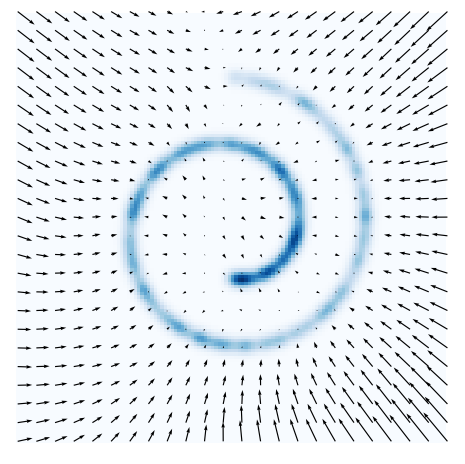} \\
    (without SN) & (with SN)
\end{tabular}
    \caption{Comparison of the score function learning with and without spectral normalization. In blue is the density of the swiss roll distribution and the black vector field is the learned score function evaluated on a grid. \nblink{SpectralNormRegularization}}
    \label{fig:spectral-norm}
\end{figure}

Sampling requires to evaluate the score function in regions the network did not necessarily observe data. As described in \autoref{sec:sampling-method}, annealing is useful to smooth the distribution, and thus the associated score. Besides, we investigated the effect of Spectral Normalization on the network regularity in those regions. Indeed, as it can be seen in figure \autoref{fig:spectral-norm}, spectral normalization smoothens the learned gradient map far from the high densities. Regularizing the spectral normal of a network lowers its Lipschitz constant, which prevents high variation between close points, thus aligning unconstrained gradients.

An important feature of the network is the noise condition. The input image in concatenated with a noise standard deviation map, as well as the input of the bottleneck of the network. We also followed the advice of \cite{Song2020} to divide the output of the network by the absolute value of the noise level. It is observed that the norm of the score function is inversely proportional to the noise power, and neural networks have issues to do this rescaling automatically. This output normalization turns out to be necessary when we consider a large order of magnitude between the noise powers during annealing.

\section{Convolution of the likelihood} \label{sec:likelihood-demo}

\textit{Proposition}: let $x \in \mathbb{R}^d$, $y \in \mathbb{R}^d$, $p_{\sigma_1}(y|x) \triangleq \mathcal{N}(y~|~\Pb x, \sigma_1^2\mathbf{I}_d)$ being the likelihood and $p_{\sigma_2}(x) \triangleq \mathcal{N}(x~|~0, \sigma_2^2\mathbf{I}_d)$ a centered multivariate Gaussian distribution. Let us assume that the operator $\Pb$ is unitary, i.e. verifies $\mathbf{P}^{\dagger} \mathbf{P} = \mathbf{I}_d$. 

Then the convolution of the likelihood with the centered gaussian is:
\begin{equation}
    p_{\sigma_1} \circledast p_{\sigma_2} (y|x) = \mathcal{N}(y~|~\Pb x, (\sigma_1^2 + \sigma_2^2)\mathbf{I}_d).
\end{equation}

\setlength\parindent{0pt} \textit{Proof}: According to the definition of $p_{\sigma_1}$ and $p_{\sigma_2}$, we have:

\medskip

$p_{\sigma_1} \circledast p_{\sigma_2} (y|x) = \int p_{\sigma_1}(y|x-t)p_{\sigma_2}(t)dt$
\begin{multline*}
  = \frac{1}{(2\pi \sigma_1^2\sigma_2^2)^{d}}\int \exp\left(-\frac{1}{2\sigma_1^2} \left( (y-\Pb(x-t))^\dagger (y-\Pb(x-t)) \right) \right) \\ 
  \times \exp\left(\frac{t^\dagger t}{2\sigma_2^2}\right) dt,
\end{multline*}

\begin{multline*}
  = \frac{1}{(2\pi \sigma_1^2\sigma_2^2)^{d}}\int \exp\Bigg(-\frac{1}{2\sigma_1^2} \bigg( \|y-\Pb x\|_2^2 + 2y^\dagger \Pb t \\ 
  - 2x^\dagger \Pb^\dagger \Pb t + t^\dagger \Pb^\dagger \Pb t \bigg) \Bigg) \exp\left(\frac{t^\dagger t}{2\sigma_2^2}\right) dt
\end{multline*}

\begin{multline*}
  = \frac{1}{(2\pi \sigma_1^2\sigma_2^2)^{d}}\int \exp\Bigg(-\frac{1}{2\sigma_1^2} \bigg( \|y-\Pb x\|_2^2 + 2y^\dagger u \\ - 2x^\dagger \Pb^\dagger u + u^\dagger u \bigg) \Bigg) \exp\left(\frac{t^\dagger t}{2\sigma_2^2}\right) dt.
\end{multline*}

We can use the change of variable $u=\Pb t$, and notice that $du = |\det \Pb| dt = dt$ and $u^\dagger u = t^\dagger \Pb^\dagger \Pb t = t^\dagger t$, since $\Pb$ is unitary, so that:

\begin{multline*}
  = \frac{1}{(2\pi \sigma_1^2\sigma_2^2)^{d}}\int \exp\Bigg(-\frac{1}{2\sigma_1^2} \left( \|y-\Pb x\|_2^2 + 2u^\dagger (y-\Pb x) + u^\dagger u \right) \Bigg) \\ \times \exp\Bigg(\frac{u^\dagger u}{2\sigma_2^2}\Bigg) du
\end{multline*}
\begin{multline*}
  = \frac{1}{(2\pi \sigma_1^2\sigma_2^2)^{d}} \exp\Bigg(-\frac{\|y-\Pb x\|_2^2}{2\sigma_1^2} \Bigg) \\
  \times \int \exp \Bigg( -\frac{u^\dagger (y-\Pb x)}{\sigma_1^2} - u^\dagger u \left( \frac{1}{2\sigma_1^2} + \frac{1}{2\sigma_2^2} \right)\Bigg) du
\end{multline*}
\begin{multline*}
  = \frac{1}{(2\pi \sigma_1^2\sigma_2^2)^{d}} \exp\Bigg(-\frac{\|y-\Pb x\|_2^2}{2\sigma_1^2} \Bigg) ~\times \\
  \int \exp \Bigg( -\frac{1}{2} \left( \frac{1}{2\sigma_1^2} + \frac{1}{2\sigma_2^2} \right) \Bigg[ \frac{2u^\dagger (y-\Pb x)}{2\sigma_1^2} \left( \frac{1}{2\sigma_1^2} + \frac{1}{2\sigma_2^2} \right)^{-1}\\ + u^\dagger u \Bigg]\Bigg) du
\end{multline*}

\begin{multline*}
  = \underbrace{\exp\Bigg(-\frac{\|y-\Pb x\|_2^2}{2\sigma_1^2} \Bigg) \exp \Bigg( \frac{\|y-\Pb x\|_2^2}{2\sigma_1^4} \left( \frac{1}{\sigma_1^2} + \frac{1}{\sigma_2^2} \right)^{-1}\Bigg)}_\text{(*)} ~\times \\
  \underbrace{\int \exp \Bigg( -\frac{1}{2} \left( \frac{1}{2\sigma_1^2} + \frac{1}{2\sigma_2^2} \right) \Bigg[ u + \underbrace{\frac{y-\Pb x}{2\sigma_1}\Bigg(\left( \frac{1}{2\sigma_1^2} + \frac{1}{2\sigma_2^2} \right)^{-1}}_{\triangleq \mu} \Bigg]^2\Bigg) du}_\text{(**)}
\end{multline*}

\begin{multline*}
  = \frac{(2\pi)^{d/2}\left(\frac{1}{\sigma_1^2} + \frac{1}{\sigma_2^2}\right)^{d/2}}{(2\pi \sigma_1^2 \sigma_2^2)^{d}} \exp \left( -\frac{\|y - \Pb x\|_2^2}{2(\sigma_1^2 + \sigma_2^2)} \right) \\
\end{multline*}
\begin{equation*}
  = (2\pi)^{-d/2}(\sigma_1^2+\sigma_2^2)^{-d/2} \exp \left( -\frac{\|y - \Pb x\|_2^2}{2(\sigma_1^2 + \sigma_2^2)} \right)
\end{equation*}

\begin{equation*}
  = (2\pi)^{-d/2}(\sigma_1^2+\sigma_2^2)^{-d/2} \exp \left( -\frac{\|y - \Pb x\|_2^2}{2(\sigma_1^2 + \sigma_2^2)} \right)
\end{equation*}

\begin{multline*}
  = \mathcal{N}(y~|~\Pb x, (\sigma_1^2 + \sigma_2^2)\mathbf{I}_d) \\
  \square
\end{multline*}

\begin{multline*}
  \text{(*)} = \exp\Bigg( -\frac{\|y-\Pb x\|_2^2}{2} \Bigg( \frac{1}{\sigma_1^2} - \frac{1}{\sigma_1^4} \Bigg( \frac{1}{\sigma_1^2} + \frac{1}{\sigma_2^2} \Bigg) \Bigg) \Bigg)\\
  = \exp \Bigg( - \frac{\|y - \Pb x\|_2^2}{2(\sigma_1^2 + \sigma_2^2)} \Bigg).
\end{multline*}

\begin{multline*}
  \text{(**)} = (2\pi)^{d/2}\left(\frac{1}{\sigma_1^2} + \frac{1}{\sigma_2^2}\right)^{d/2} ~\times\\
  \underbrace{\frac{1}{(2\pi)^{d/2}\left(\frac{1}{\sigma_1^2} + \frac{1}{\sigma_2^2}\right)^{d/2}} \int \exp \left( -\frac{1}{2}\left( \frac{1}{\sigma_1^2} + \frac{1}{\sigma_2^2} \right) [u + \mu] \right)du}_{=1}.
\end{multline*}

\section{Signal-to-Noise Ratio of the convergence field $\kp$} \label{sec:snr}

\begin{figure}[h]
    \centering
    \includegraphics[width=\linewidth]{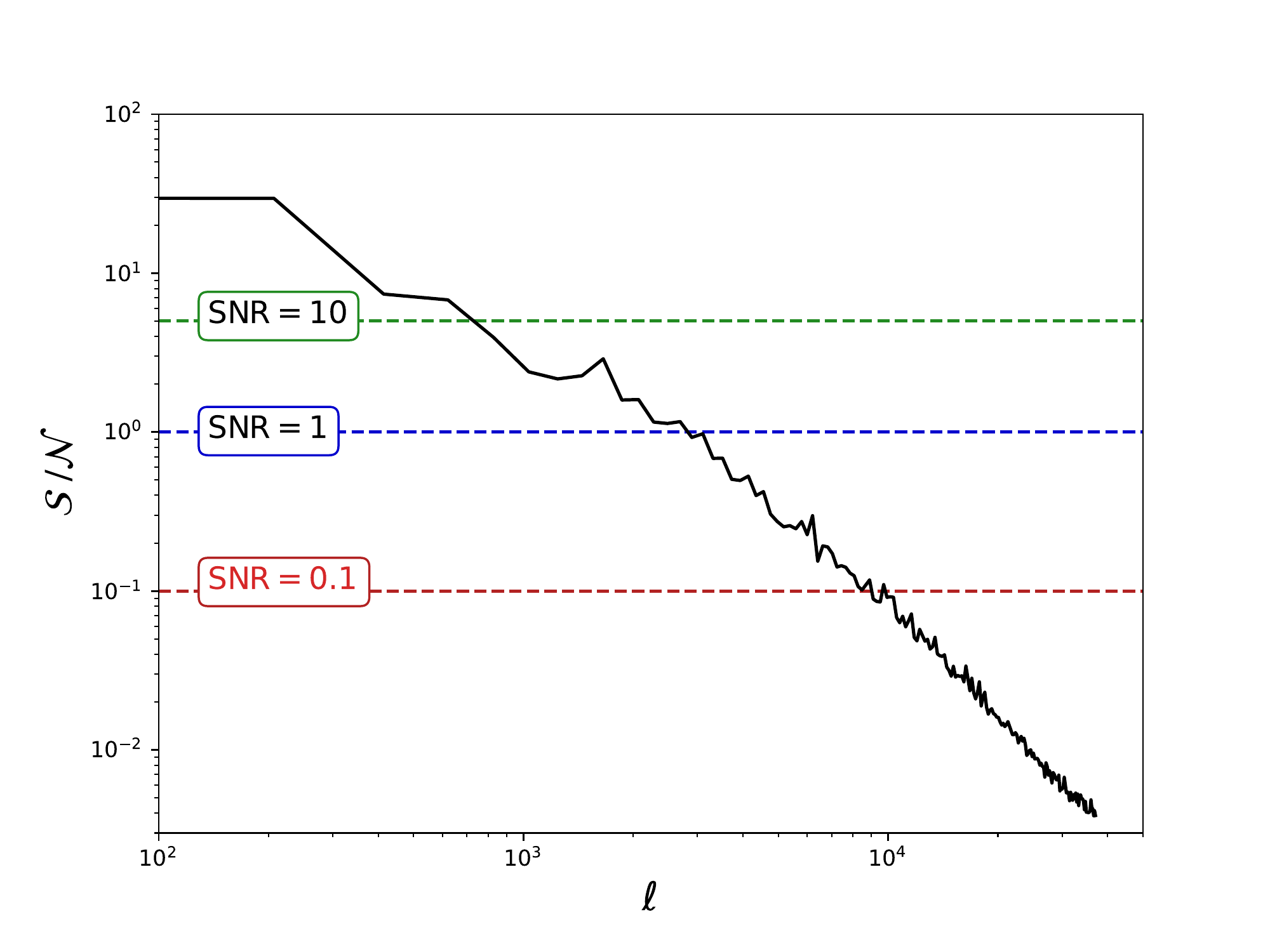}
    \caption{SNR of the input data. The SNR corresponds to the ratio between the fiducial power spectrum and the power spectrum of the input noise. This shows that the SNR being equal to 0.1 from $\ell=10^4$, therefore the reconstruction from this scale does not corresponds to the input data, but is driven by the prior only. \nblink{WienerGaussianPrior}}
    \label{fig:snr-input}
\end{figure}

In figure \autoref{fig:snr-input} we show the ratio between the fiducial power spectrum and the power spectrum of the input noise.

\section{Cluster detection}

\begin{figure}[h!]
    \centering
    \begin{tabular}{c}
    Detection \\
    \includegraphics[width=\linewidth]{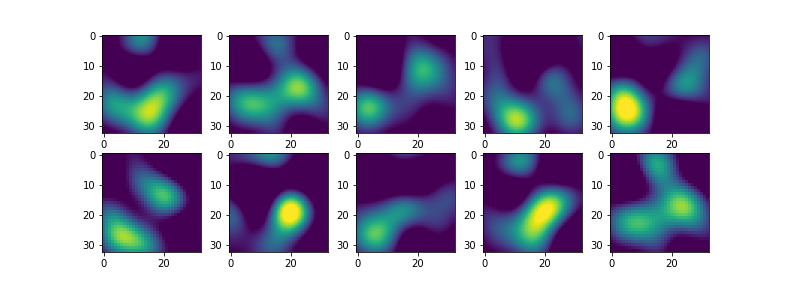}\\
    No detection \\
    \includegraphics[width=\linewidth]{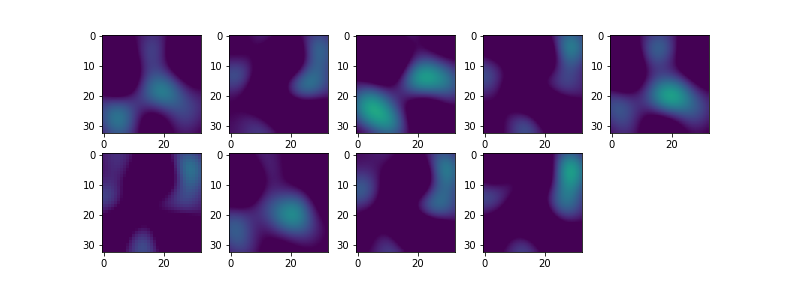}
    \end{tabular}
    \caption{This figure shows filtered maps of the convergence posterior samples estimated by our method. In the upper panel, a cluster was added in the input shear, while not below. Above cutouts of the filtered maps show the recovered cluster shape in the center of the map, while below cutouts show coefficient from structure nearby. Selection was done with a 3$\sigma$ threshold. \nblink{Detection}}
    \label{fig:stamps_detection}
    
\end{figure}

In the detection experiment described in \autoref{sec:full-prior-detection}, we added a NFW profile into the input shear map. We the ran our cluster detection procedure on the associated \texttt{DLPosterior} samples. \autoref{fig:stamps_detection} shows cutouts of those samples around the cluster location. The upper plot shows the samples that were selected with a $3\sigma$ threshold and the bottom plot the ones below. We can clearly recognize the shape of a cluster among the samples with positive detection, i.e. with maximum coefficient above $3\sigma$.

\section{Mass-Mapping methods comparison. } \label{sec:map-comparison}
\begin{figure*}
    \centering
    \includegraphics[width=\textwidth]{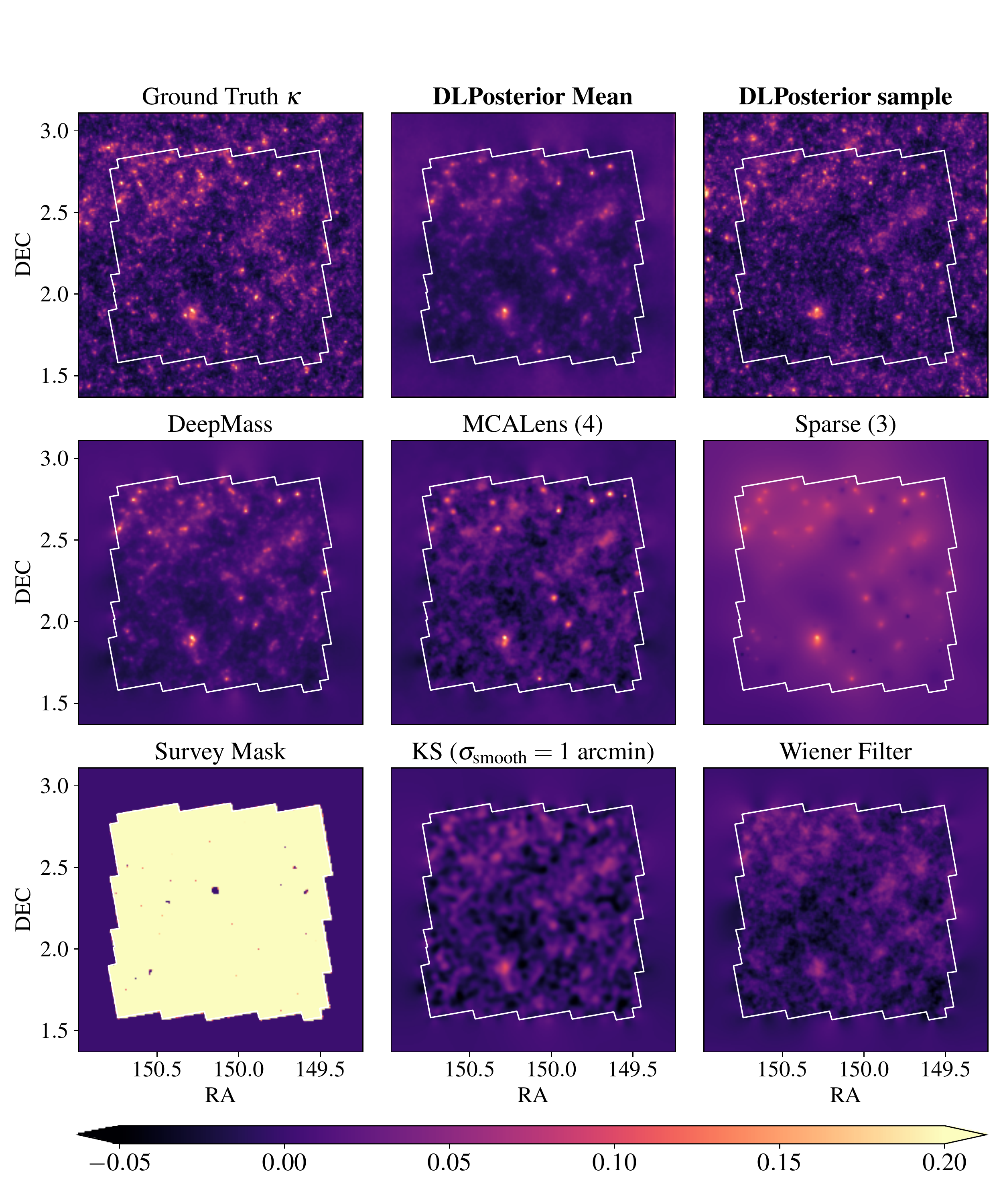}
    \caption{Mass-mapping methods comparison. \textit{Ground truth} corresponds to the convergence map $\kappa$ that we aim to recover, taken from \citet{Osato2021}, \texttt{DLPosterior} mean and \texttt{DLPosterior} sample are the results discussed in \autoref{sec:validation}, \texttt{DeepMass} is from \citet{Jeffrey2020}, \texttt{MCALens} with $\lambda=4$ is from \citet{sta:starck21}, \texttt{GLIMPSE} method with $\lambda=3$ is from \citet{Lanusse2016}, \textit{Survey Mask} is the COSMOS catalog binary mask, \textit{Kaiser-Squires} with Gaussian smoothing ($\sigma_\text{smooth}=1$ arcmin) is from \citet{Kaiser1993} and the \textit{Wiener Filter} is from \citet{elsner2013, Jeffrey2018}. \nblink{MassMappingResults}}
    \label{fig:method-comparison}
\end{figure*}

In figure \ref{fig:method-comparison}, we show how the different methods described in \autoref{sec:massmapping} recover a convergence map from the same input shear field used in \autoref{sec:validation} and COSMOS-like survey mask.

\end{appendix}

\end{document}